\documentclass[12pt,english,preprint]{article}
\usepackage[T1]{fontenc}
\usepackage[latin9]{inputenc}
\usepackage[a4paper]{geometry}
\usepackage{float}
\usepackage{amsthm}
\usepackage{amsmath}
\usepackage{amssymb}
\usepackage{graphicx}
\usepackage{setspace}
\usepackage{wasysym}
\usepackage{esint}
\usepackage{babel}
\begin{document}

\title{Numerical study of the gravitational shock wave inside a spherical
charged black hole }

\author{Ehud Eilon and Amos Ori}

\maketitle
\noindent \begin{center}
\emph{Department of Physics, }
\par\end{center}

\noindent \begin{center}
\emph{Technion - Israel Institute of Technology, }
\par\end{center}

\noindent \begin{center}
\emph{Haifa 3200003, Israel}
\par\end{center}
\begin{abstract}
We numerically investigate the interior of a four-dimensional, asymptotically
flat, spherically symmetric charged black hole perturbed by a scalar
field $\Phi$. Previous study by Marolf and Ori indicated that late
infalling observers will encounter an effective shock wave as they
approach the left portion of the inner horizon. This shock manifests
itself as a sudden change in the values of various fields, within
a tremendously short interval of proper time $\tau$ of the infalling
observers. We confirm this prediction numerically for both test and
self-gravitating scalar field perturbations. In both cases we demonstrate
the effective shock in the scalar field by exploring $\Phi(\tau)$
along a family of infalling timelike geodesics. In the self-gravitating
case we also demonstrate the shock in the area coordinate $r$ by
exploring $r(\tau)$. We confirm the theoretical prediction concerning
the shock sharpening rate, which is exponential in the time of infall
into the black hole. In addition we numerically probe the early stages
of shock formation. We also employ a family of null (rather than timelike)
ingoing geodesics to probe the shock in $r$. We use a finite-difference
numerical code with double-null coordinates combined with a recently
developed adaptive gauge method in order to solve the (Einstein +
scalar) field equations and to evolve the spacetime (and scalar field)
--- from the region outside the black hole down to the vicinity of
the Cauchy horizon and the spacelike $r=0$ singularity.

\newpage{}
\end{abstract}

\section{INTRODUCTION \label{sec:introduction}}

\paragraph*{Background: }

The inner structure of classical black holes (BHs) has been a subject
of continuous investigation over the last half century. The interior
of a Reissner-Nordström (RN) or Kerr BH (representing a charged or
spinning BH respectively) is drastically different from that of a
Schwarzschild BH. In the latter there is a fatal, destructive, spacelike
singularity at $r=0$, and all infalling observers inevitably crash
at that singularity. By contrast, in the RN and Kerr solutions there
is an inner horizon (IH) at finite $r$, instead of a spacelike singularity.
An $r=0$ singularity still exists in these two solutions, but it
is timelike rather than spacelike, it is located beyond the inner
horizon, and infalling geodesics generically avoid it (unlike the
Schwarzschild case). Figure 1a displays the Penrose diagram of the
eternal (analytically extended) RN spacetime. The Kerr case is basically
similar (at least with regards to the aspects considered in this paper).
\footnote{ Certain differences and additional subtleties arise in the spacetime
diagram of Kerr, but they all occur beyond the IH and are hence not
so relevant to the present paper. (To mention two such differences:
The $r=0$ singularity in Kerr is actually a \emph{ring}, and at its
``other side(s)'' there is an additional external asymptotic universe,
of a different type.) } 

The IH in the RN and Kerr spacetimes also serves as a \emph{Cauchy
horizon} (CH), a null hypersurface which marks the future boundary
of physical predictability (for initial data specified in the external
world). Note that in an RN or Kerr BH there are two distinct inner-horizon
hypersurfaces (see Fig. 1a), namely a left-going arm and a right-going
one. In the case of eternal BH, the CH contains both IH arms (up to
their bifurcation point b). This is directly related to the presence
of two external asymptotic universes, W and W'. However, in the case
of a non-eternal charged BH produced in spherical charged collapse
(see Fig. 1b), the left-right symmetry breaks, and the asymptotic
universe W' no longer exists. In this case only the left-going arm
of the IH is a CH. The other IH arm seems to have no special causal
role in the charged-BH spacetime. The situation with spinning BHs
is basically similar: In the eternal Kerr spacetime the CH contains
both arms of the IH; but in the non-eternal, spinning-collapse case
(with asymptotically-Kerr exterior) only the left-going arm belongs
to the CH. \footnote{In the case of spherical charged collapse, the spherically-symmetric
electrovac exterior will uniquely be described by the RN geometry.
But in the spinning analog, the vacuum exterior will generically fail
to be Kerr. Indeed, at late time the geometry should approach a Kerr
BH (the ``no hair'' principle). However, in this case the CH will
become a weak curvature singularity (in contrast with the regular
CH in the cases of pure Kerr and RN geometries, and in the case of
unperturbed spherical charged collapse.) We shortly discuss this type
of weakly-singular CH. }

\begin{center}
\begin{figure}[H]
\begin{centering}
\includegraphics[scale=0.75]{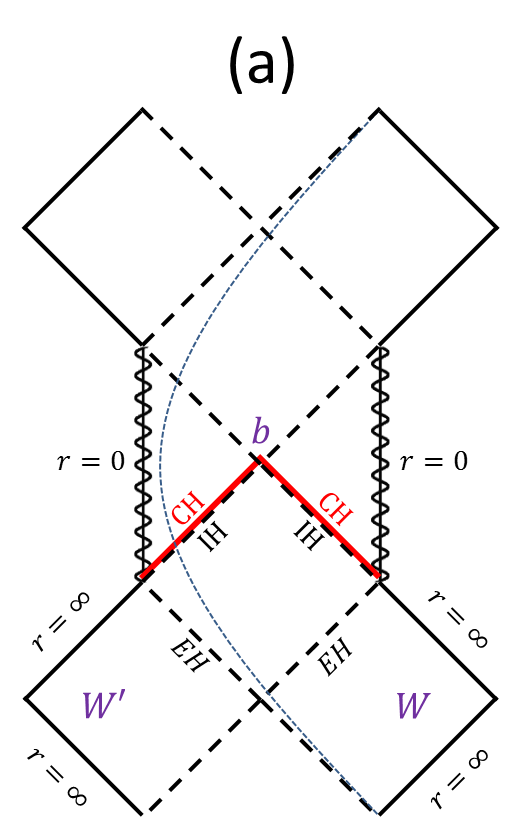}\hspace{3cm}\includegraphics[scale=0.75]{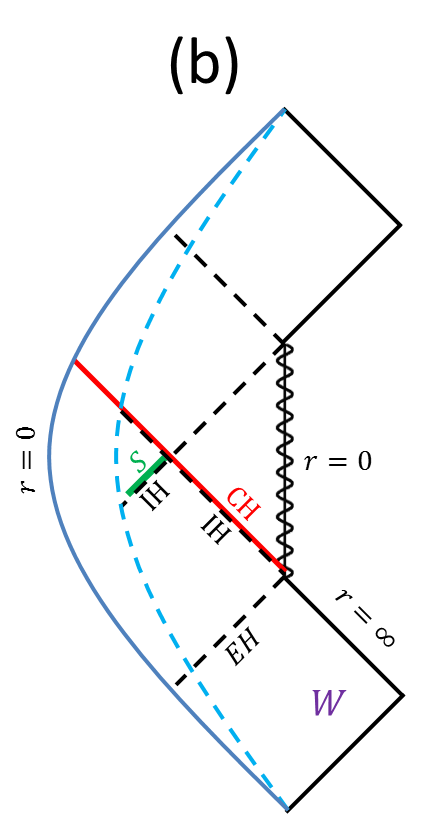}
\par\end{centering}

\protect\caption{\label{fig:Internal-structure} Penrose diagrams of the eternal RN
spacetime (panel a) and a charged BH produced by thin-shell collapse
(panel b). In the latter, the geometry is RN outside the collapsing
shell (the dashed blue curve) and Minkowski inside the shell, with
a regular center at $r=0$ (the solid blue curve). In both panels,
solid black diagonal lines denote null infinity; solid red diagonal
lines denote the Cauchy horizon (CH). Dashed black diagonal lines
denote the event horizon (EH) and the inner horizon (IH); the latter
partly overlaps with the CH in both cases. Wavy vertical lines denote
the timelike $r=0$ singularity of RN. The eternal RN spacetime in
panel a includes a twin set of asymptotically flat universes, W and
W'. This panel also shows a typical timelike geodesic (the dotted
blue curve) which falls into the BH from the external universe W,
avoids the timelike $r=0$ singularity, then emerges from a white
hole and arrives at a future external universe (isometric to W). The
spacetime diagram in panel a is right-left symmetric; and the CH has
two arms which intersect at a bifurcation point ``b''. The charged-collapse
diagram in panel b is not right-left symmetric, and the CH includes
only one arm of the IH (the left-going one). When the collapsing charged
BH of panel b is generically perturbed, a null week singularity forms
at the CH; and an effective shock develops along the solid green line
denoted ``S'' (at the right-going IH).}
\end{figure}

\par\end{center}

In the late 1960s Penrose pointed out that the CH, in both RN and
Kerr geometries, is a locus of infinite blue-shift. \cite{1968-Penrose-PRN-Analytical}
He suggested that this diverging blue-shift would lead to divergent
energy fluxes, and hence to a curvature singularity instead of a regular
IH. Since then, many investigations were made in an attempt to analyze
the effect of perturbations on the internal structure of the (charged
or spinning) BH --- and particularly at the CH. In order to simplify
the analysis, most of these investigations \cite{1973-Simpson-Penrose-PRN-Numerical,1981-Hiscock-PRN-Analytical,Novikov,1982-Chand-Hartle-PRN-Analytical,Poisson-Israel-PRL,1990-Poisson-Israel-mass-function,Ori-Weak-singularity,1993-Gnedin,1995-Brady-Smith,1997-Burko-Internal-structure,1998-Hod-Piran,Burko-EOMs}
were restricted to spherical charged BHs, because spherical symmetry
drastically simplifies the structure of spacetime. Of particular importance
were the works of Hiscock \cite{1981-Hiscock-PRN-Analytical} and
of Poisson and Israel \cite{Poisson-Israel-PRL,1990-Poisson-Israel-mass-function}:
Hiscock considered an ingoing null fluid perturbation, and showed
that a (non-scalar) null curvature singularity forms at the CH. Poisson
and Israel introduced the mass-inflation model, which contains two
null fluids. They concluded that a scalar-curvature null singularity
will form at the CH in this case, the so-called \emph{mass-inflation
singularity}. Ori \cite{Ori-Weak-singularity} then found (using a
simplified shell model) that the mass-inflation singularity is actually
weak (in Tipler's \cite{Tipler} sense; see also \cite{Ori-What-is-Weak}),
despite of its scalar-curvature character. Namely, the metric tensor
approaches a finite (non-singular) limit at the CH; and extended objects
will only experience finite (possibly small) tidal distortion on approaching
the CH. Subsequently, several authors extended these studies to the
case of a spherical charged BH perturbed by a self-gravitating scalar
field. Numerical investigations \cite{1995-Brady-Smith,1997-Burko-Internal-structure}
revealed that in this case too, a weak null curvature singularity
develops at the CH. In addition, the CH undergoes contraction due
to crossing energy fluxes, and eventually it shrinks to zero area
(namely $r=0$). Various numerical simulations \cite{1995-Brady-Smith,1997-Burko-Internal-structure,1998-Hod-Piran,Numerical-Methods}
indicated that at this point of full contraction, the null CH meets
a spacelike $r=0$ singularity, as seen in Fig. \ref{fig:NL-num-grid}
below. This basic picture, of a weak null singularity forming at the
CH, was recently confirmed mathematically by Dafermos \cite{Dafermos}. 

The non-spherical model of a rotating BH is of course much more realistic
than its spherical-charged counterpart (as astrophysical BHs are presumably
rotating but not significantly charged). Considering the case of a
perturbed asymptotically-flat spinning BH, it was found \cite{1992-Ori-spinning}
(based on nonlinear perturbation analysis) that the situation is similar
in many respects to the case of perturbed spherical charged BHs discussed
above: Owing to the infinite blue-shift of the infalling gravitational
perturbations, a null weak scalar-curvature singularity develops at
the CH in this case too. Later an independent analysis \cite{Brady-Droz}
yielded consistent results. Note a remarkable difference between the
spinning and spherical-charged cases, though: In the spinning case
the CH singularity turns out to be oscillatory \cite{2000-Ori-Spinning},
unlike the monotonic growth of curvature scalars in the mass-inflation
singularity. 

The accumulation of all these investigations, over the last few decades,
led to a fairly clear and coherent picture: \footnote{We point out, however, that several cardinal questions concerning
the interior of classical BHs still remain open. Perhaps the most
important one is, whether a spacelike singularity forms inside a generically-perturbed
spinning BH.} Consider either a spinning or charged BH produced in gravitational
collapse. A (null and weak) scalar-curvature singularity will generically
develop at the CH, namely the \emph{left-going} section of the IH.
No such curvature singularity is expected to form at the latter's
right-going section --- which, as mentioned above, is \emph{not} a
CH. \cite{2010-Hamilton-Avelino}

\paragraph*{Experience of infalling observers:}

Consider now a hypothetical observer who arrives a charged (or spinning)
non-eternal BH a long time after the collapse, and decides to jump
in. In view of the no-hair principle, the BH will look pretty much
like RN (or Kerr), with only negligible deviations, perhaps even undetectable
ones. This asymptotic RN metric will have a final mass and charge
parameters $M_{f}$ and $Q_{f}$, and hence outer and inner horizon
radius $r_{\pm}=M_{f}\pm\left(M_{f}^{2}-Q_{f}^{2}\right)^{1/2}$,
where relativistic units $c=G=1$ are used throughout the paper. (In
the Kerr case, $Q_{f}$ should be replaced by the asymptotic spin
parameter $a_{f}$.) Once inside the BH, the observer will inevitably
fall towards the IH, and its $r$ value will monotonically decrease
from $r_{+}$ to $r_{-}$. Throughout most of this journey from the
EH to the IH, the deviations from the RN/Kerr metric are still negligible
--- and, ignoring these tiny deviations, the observer would have the
same experience as moving in an exact RN/Kerr geometry. Assuming that
the observer is equipped with a rocket, he can choose whether to reach
$r=r_{-}$ at the right portion of the IH, or at its left portion.
We shall refer to such observers as \emph{right-fallers} or \emph{left-fallers},
respectively. \footnote{To avoid confusions that may potentially be associated with this ``right-left''
terminology, we emphasize that the \emph{right} (or left) portion
of $r=r_{-}$ is \emph{left}-going (or right-going); See Fig. 1a.
Note also that in the RN case, an infalling geodesic will always take
the observer directly to the left portion of $r=r_{-}$; he would
have to turn on his rocket in order to reach the right portion. However
in the Kerr case geodesics reach either the left or right portions.} 

Let's consider first the experience of a right-faller. The right portion
of $r_{-}$ is the CH, which, as was discussed above, is a locus of
a curvature singularity. Throughout most of the travel from $r_{+}$
to $r_{-}$ the observer will feel nothing but the background RN/Kerr
curvature (which would be fairly mild for a supermassive BH), with
only negligible deviations. Only when $r$ becomes very close to $r_{-},$
tidal forces will start to grow very rapidly, and will diverge at
the CH itself. (Although, the integrated deformation will be finite
even at the CH itself \footnote{Recall that, since the CH singularity is weak, the integrated tidal
deformation is finite (and non-destructive) even at the singularity;
furthermore, for sufficiently late infall time, the overall singularity-induced
deformation may be arbitrarily small. Nevertheless, the tidal force
itself will always diverge at the singular CH. }.) This is the basic picture that emerges from the collection of the
various investigations over the last few decades. 

But what will be the experience of a left-faller? The above-mentioned
investigations haven't directly addressed this question, except that
they collectively made clear one important fact: Since the BH is non-eternal
(it has presumably been produced in gravitational collapse), the left
portion of $r=r_{-}$ is \emph{not} a locus of divergent curvature
(because it is not a CH); hence, in principle spacetime should be
perfectly smooth on crossing this portion of the IH. 

Of course, the left-faller should expect very serious troubles to
come later on; for example, a strong spacelike singularity may exist
in the perturbed spacetime, say at $r=0$. But this should happen
somewhere at $r<r_{-}$, namely sometime \emph{after} crossing $r=r_{-}$.
Exactly where the fatal tidal forces will be encountered, this might
depend on the details of the BH model. These considerations might
lead to the impression that the experience of a left-faller is rather
non-universal and hard to predict, and in particular it might be sensitive
to various details of the collapse scenario.

\paragraph*{Effective shock wave:}

Despite of the above, Marolf and Ori (MO) \cite{Marolf-Ori_Shockwave}
recently found that the experience of late left-fallers, when recorded
as a function of their own proper time $\tau$, does exhibit an interesting
universal pattern --- a shock-like behavior. Namely, various measurable
quantities undergo abrupt changes as $r=r_{-}$ is approached. For
any individual left-faller, this change actually takes a finite lapse
of proper time $\Delta\tau$ (in line with the aforementioned regularity
of spacetime at the left section of $r=r_{-}$). However, this finite
width $\Delta\tau$ rapidly decreases with delaying the moment of
infall into the BH. Based on simple theoretical arguments, MO concluded
that $\Delta\tau\propto e^{-\kappa_{-}\Delta t}$, where $\kappa_{-}$
is the IH surface gravity, and $\Delta t$ is (roughly speaking) the
time lapse from BH formation to the moment of jumping in, expressed
in terms of asymptotic time coordinate $t$ in the external universe.
(Later we shall re-express this exponential relation more precisely
in terms of Eddington's advanced time coordinate $v_{e}$.) For a
left-faller which jumps in after a time delay $\Delta t$ of, say,
a few tens or hundreds times the BH mass $M$, the width scale $\Delta\tau$
will be tremendously smaller than Planck time, by many orders of magnitude;
so small that it is not clear if it could be resolved by any physical
probe. We therefore refer to this rapid change in the various measurable
quantities as an ``effective discontinuity'', or \emph{effective
shock wave}, which forms at the left-arm IH. 

The analysis by MO addressed both charged and spinning BHs. The effective
shock wave shows itself in various quantities, e.g. metric functions
as well as other perturbing fields. In this paper we consider the
model of a spherical charged BH perturbed by a scalar field, and focus
on the shock formation in two quantities: the perturbing scalar field,
and the area coordinate $r$ --- which is essentially a (square root
of the) metric function for the angular sector.

\paragraph*{Numerical verification:}

The analysis in Ref. \cite{Marolf-Ori_Shockwave} was carried entirely
analytically, by focusing on the late-time behavior of the relevant
perturbation fields inside the BH. The investigation was thus heavily
based on perturbation analysis, although in certain points it also
involved some qualitative considerations concerning the dynamics beyond
the small-perturbation domain. 

As an example, consider the shock in $r$: It involves a sudden decrease
of $r$ from $r_{-}$ towards $r=0$, within an extremely small $\Delta\tau$
(see Sec. \ref{subsec:shock-in-r} below). This clearly involves a
large deviation of $r(\tau)$ from its smooth RN counterpart. This
situation is thus beyond the domain of validity of perturbation analysis.
Hence it would be worthwhile to confirm this shock behavior by independent,
non-perturbative, analysis. Another limitation of the analysis by
MO was that it was mostly restricted to the late-time domain, and
therefore was incapable of resolving the earlier times where the shock
just starts to develop (namely, the domain of small-moderate $\Delta t$
values at which $\Delta\tau$ starts its decrease from its initial
$\mathcal{O}(M)$ value to its later $\ll M$ values). 

We therefore found it important to carry detailed numerical simulations
for exploring the shock phenomenon inside the BH. Specifically we
were motivated by several goals and objectives, including: (i) verifying
the very existence of the effective shock, (ii) verifying the shock's
exponential sharpening rate, (iii) exploring the early phase of shock
formation, and (iv) exploring the well-developed shock in $r$, all
the way down to $r\ll r_{-}$ values (which, as pointed out above,
are out of the perturbative domain). 

To this end, we carried a numerical analysis of a spherical charged
BH perturbed by either a test scalar field or a self-gravitating one.
We used the double-null numerical code \cite{Numerical-Methods} which
we recently developed for that purpose. We study the effective shock
in the scalar field and (in the self-gravitating case) also in the
area coordinate $r$. We explore both the early shock formation and
the later domain of fully developed shock. In the latter domain, we
also measure the rate of shock sharpening, and compare it to MO's
prediction in \cite{Marolf-Ori_Shockwave}. We found nice agreement
between this prediction and our numerical results, concerning both
the very existence of the shock and its rate of sharpening. 

The paper is organized as follows: We formulate the physical problem
in terms of the unknown functions, the field equations, and the set-up
of initial conditions in Sec. \ref{sec:physical-system}; we summarize
our numerical algorithm in Sec. \ref{sec:Basic-Numerical-Algorithm}.
Since we discussed both of these subjects extensively in our previous
paper \cite{Numerical-Methods}, here we only summarize them briefly.
We then numerically analyze the case of a test scalar field perturbation
on a prescribed RN background in Sec. \ref{sec:test-scalar-field}.
In particular, we demonstrate the formation of effective shock in
the scalar field, and show that its development exhibits a continuous
sharpening, in nice quantitative agreement with the prediction made
by MO. The case of a self-gravitating scalar field perturbation on
a spherical charged background is analyzed in Sec. \ref{sec:self-gravitating-scalar-field},
where we demonstrate the expected shock formation in both the scalar
field and the metric function $r$. The rate of shock sharpening (for
both the scalar field and $r$) is measured and compared to MO's prediction.
In addition, we also explore the shock in $r$ as expressed in terms
of the affine parameter $\lambda$ along null (rather than timelike)
ingoing geodesics. We find that null geodesics provide an efficient
tool for exploring the shock. Finally, we discuss our results and
conclusions in Sec. \ref{sec:discussion}.

\section{PHYSICAL SYSTEM AND FIELD EQUATIONS \label{sec:physical-system}}

We explore the formation and development of the shock wave in two
different physical scenarios: (i) Evolving \emph{test scalar field}
on a prescribed, static, RN background; and (ii) the evolution of
a \emph{self-gravitating scalar field} on a dynamical charged BH background. 

In the first scenario the background metric is fully known, a-priori
and analytically; in principle one could solve the scalar-field wave
equation only, on the prescribed RN background. We found it more convenient,
however, to numerically solve the Einstein equation as well --- with
electrovac initial data. The evolving metric then yields the RN solution
(up to negligible numerical errors). This set-up allows more flexible
switching between scenarios (i) and (ii), which in particular yields
better debugging capabilities. Also, it allows easier handling of
the gauge used for the background RN geometry --- simply by dictating
the gauge of the initial functions. As a consequence, the basic numerical
setting of the self-gravitating system described in this section ---
namely simultaneous numerical solution of the Einstein and scalar-field
equations --- actually applies to the test-field case as well. \footnote{For similar reasons of numerical-scheme uniformity, we also chose
to numerically treat the test scalar field as a self-gravitating one
but with an ``effectively infinitesimal'' pre-factor, which we took
here to be $A_{0}=10^{-20}$. (The self-gravity effects are then all
multiplied by $10^{-40}$, which is smaller than our numerical resolution
by many orders of magnitude.) \label{fn: test-field}}

The scalar field $\Phi$ is uncharged, massless, and minimally coupled,
satisfying the standard wave equation $\Square\Phi=0$. The geometry
is initially RN, but is subsequently perturbed by an ingoing pulse
of scalar field (see below). The initial RN geometry has mass $M_{0}$
and charge $Q$. The line element may be expressed in double-null
coordinates $(u,v,\theta,\varphi)$ (applicable to both the RN and
perturbed geometries):

\begin{equation}
ds^{2}=-e^{\sigma(u,v)}dudv+r(u,v)^{2}d\Omega^{2}\,,\label{eq:line-element}
\end{equation}

\noindent where $d\Omega^{2}\equiv d\theta^{2}+\sin^{2}\theta\,d\varphi^{2}$.
Our three unknown functions are the scalar field $\Phi(u,v)$ and
the metric functions $r(u,v)$ and $\sigma(u,v)$.

The general setup of our simulation, and the location of the numerical
grid in the (would-be) RN background, are illustrated in Fig. \ref{fig:RN-num-grid}.
The scalar field perturbation is taken to be an ingoing initial pulse
which propagates toward the BH, and is partly scattered and partly
absorbed by it. The initial pulse has a finite support on the outgoing
initial ray $u=u_{0}$, it starts at $v=v_{1}$ and ends at $v=v_{2}$.
Correspondingly, at $v<v_{1}$ the geometry is precisely RN with mass
$M_{0}$ and charge $Q$. In the test field case, this situation remains
unchanged throughout the simulation. In the self-gravitating case,
the metric is no longer RN (and no longer static) at $v>v_{1}$. In
particular the mass function (defined below) steadily grows with $v$. 

\begin{center}
\begin{figure}[H]
\begin{centering}
\includegraphics[scale=0.6]{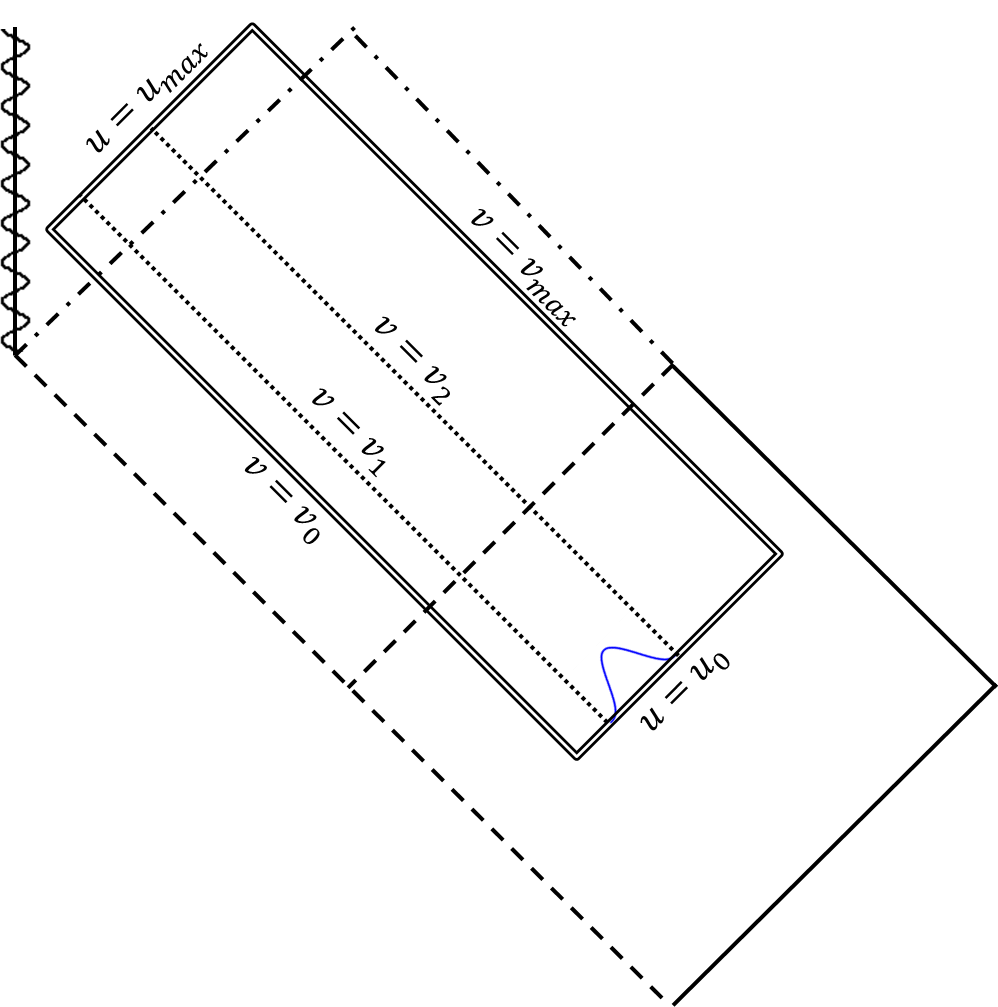}
\par\end{centering}

\protect\caption{\label{fig:RN-num-grid}This figure describes the location of the
numerical domain of integration with respect to the (initial) RN spacetime.
Double lines represent the boundary of numerical domain $(u=u_{0},\,u=u_{max},\,v=v_{0}$
and $v=v_{max}$); Solid lines represent future and past null infinity
$(r=\infty)$. Dashed and dashed-dotted lines represent the event
horizon $(r=r_{+})$ and inner horizon $(r=r_{-})$ respectively.
The ingoing scalar field pulse is drawn schematically on the outgoing
initial ray $u=u_{0}$. The pulse is confined (on $u=u_{0}$) to the
range $v_{1}\leq v\leq v_{2}$. The null rays $v=v_{1}$ and $v=v_{2}$
(the putative pulse boundaries) are marked by dotted lines.}
\end{figure}

\par\end{center}

Two remarks should be made here: First, in Fig. 2 we placed the rectangular
numerical domain of integration on the eternal RN background, just
for simplicity. However, we can equally well place this rectangle
in the RN exterior of a collapsing charged shell, shown in Fig. 1b.
In fact, this latter setting is the more physically motivated one,
because the physical situation mostly relevant to shock formation
is that of charged (or spinning) BH produced in gravitational collapse.
The same remark also applies to the self-gravitating set-up illustrated
in Fig. \ref{fig:NL-num-grid}. Second, ideally we would like to explore
the evolution of perturbations up to the CH, located at $v\to\infty$.
For obvious practical reasons we have to pick a finite $v_{max}$
value in our numerics. Nevertheless, below we shall choose a sufficiently
large $v_{max}$, such that the domain of integration will extend
deeply into the late-time domain, and will expose the evolution of
the various perturbation fields up to the neighborhood of the CH. 

The derivation of the model's field equations is fairly standard,
it is described in some detail in our preceding paper \cite{Numerical-Methods}
(see in particular Sec. 2), as well as in other \cite{Burko-EOMs}
previous works. Overall, we have a system of three evolution equations
\begin{equation}
\Phi,_{uv}=-\frac{1}{r}(r,_{u}\Phi,_{v}+r,_{v}\Phi,_{u})\,,\label{eq:  phi_evolutio}
\end{equation}

\noindent 
\begin{equation}
r,_{uv}=-\frac{r,_{u}r,_{v}}{r}-\frac{e^{\sigma}}{4r}(1-\frac{Q^{2}}{r^{2}})\,,\label{eq:  r_evolutio}
\end{equation}
\begin{equation}
\sigma,_{uv}=\frac{2r,_{u}r,_{v}}{r^{2}}+\frac{e^{\sigma}}{2r^{2}}(1-\frac{2Q^{2}}{r^{2}})-2\Phi,_{u}\Phi_{,v}\,\,,\label{eq:  sigma_evolutio}
\end{equation}
and two constraint equations 
\begin{equation}
r,_{uu}-r,_{u}\sigma,_{u}+r(\Phi,_{u})^{2}=0\,,\label{eq:  ruu}
\end{equation}
\begin{equation}
r,_{vv}-r,_{v}\sigma,_{v}+r(\Phi,_{v})^{2}=0\,.\label{eq:  rvv}
\end{equation}
The constraint equations need only be imposed at the initial hypersurface,
due to the consistency of the evolution and constraint equations. 

We shall occasionally use the \emph{mass function} $m(u,v)$, introduced
in Ref. \cite{1990-Poisson-Israel-mass-function}, which translates
in our coordinates to

\noindent 
\begin{equation}
m=(1+4e^{-\sigma}r,_{u}r,_{v})r/2+Q^{2}/2r\,.\label{eq:mass_formula}
\end{equation}

The location of the event horizon is important in various aspects
of this investigation. Here we define it to be the (first) $u$ value
where $r_{,v}$, evaluated at the final ingoing ray of the numerical
grid $v=v_{max}$, changes its sign from positive to negative. We
denote it as $u_{h}$. We define the black-hole ``final mass'' $m_{final}$
as the value of the mass function at the intersection point $(u_{h},v_{max})$.
\footnote{Note that, owing to the finiteness of $v_{max}$, the true horizon
location is slightly earlier than $u_{h}$. Also, the final BH mass
is slightly larger than our $m_{final}$. Both effects are caused
by scalar-field inward back-scattering that takes place at $v>v_{max}$.
These two deviations are negligibly small, however, owing to asymptotic
flatness combined with the large values of $v_{max}$ (always taken
to be $\gg m_{final}$) in our simulations. See also footnote \ref{fn:m_final_drift}.} In the test field case, $m_{final}=M_{0}$ of course. 

The characteristic initial hypersurface consists of the two null rays
$u=u_{0}$ and $v=v_{0}$ (below we set $u_{0}=v_{0}=0$ in the numerics).
On each initial ray we choose initial conditions for two functions,
$\Phi$ and $\sigma$. The remaining function $r$ is dictated {[}apart
from initial conditions at the vertex ($u_{0},v_{0}$){]} by the relevant
constraint equation --- Eq. (\ref{eq:  ruu}) at $v=v_{0}$ and Eq.
(\ref{eq:  rvv}) at $u=u_{0}$. The choice of initial conditions
for $\sigma$ amounts to a gauge choice: under the gauge transformation
$v\rightarrow v'(v),u\rightarrow u'(u)$, the variables $r$ and $\Phi$
are unchanged, while $\sigma$ changes according to
\begin{equation}
\sigma\rightarrow\sigma'=\sigma-\ln(\frac{du'}{du})-\ln(\frac{dv'}{dv})\,\,.\label{eq:  sigma_gauge}
\end{equation}
The gauge is thus determined by the choice of initial conditions for
$\sigma$ along $u=u_{0}$ and $v=v_{0}$.

\section{BASIC NUMERICAL ALGORITHM\label{sec:Basic-Numerical-Algorithm}}

We discreticize the field equations on a double-null grid with fixed
spacing $\Delta u,\Delta v$. Usually we take $\Delta u=\Delta v=\frac{M_{0}}{N}$,
where $M_{0}$ is the initial BH mass and $N$ takes several values
in each run (typically $80,160,320,640$), in order to verify numerical
convergence. The numerical solution progresses along rays of constant
$u$, starting from the initial ray $u=u_{0}$ and up to the final
ray $u=u_{max}$; along each outgoing ray the solution is advanced
step by step from $v=v_{0}$ to $v=v_{max}$. We discreticize the
evolution equations (\ref{eq:  phi_evolutio}-\ref{eq:  sigma_evolutio})
and apply a predictor-corrector scheme, with second order accuracy,
as described in Ref. \cite{Numerical-Methods}. 

This basic scheme works very well (second order convergence in all
unknowns) as long as the domain of integration does not penetrate
into the BH. Even if it does penetrate, it still works well as long
as $\Delta v_{e}$, the grid size (in outgoing direction) in terms
of advanced Eddington coordinate $v_{e}$ (defined below), is not
too large. However, when the horizon is included and $\Delta v_{e}\gg M_{0}$,
the numerical error typically grows as $e^{\kappa_{+}v_{e}}$ along
the EH, where $\kappa_{+}$ is the latter's surface gravity. If $\Delta v_{e}$
is grater than, say, $20$ or $25$ times $M_{0}$ (the actual number
depends on $\kappa_{+}$ and on a few other numerical parameters),
the truncation error runs out of control and the numerics breaks down.
This phenomenon was demonstrated and thoroughly analyzed in Ref. \cite{Numerical-Methods}.

\subsection{The maximal-$\sigma$ gauge \label{sub:The-maximal--gauge}}

In order to circumvent this problem of exponentially-growing truncation
error, we use the \emph{maximal-$\sigma$ gauge} (see Sec. 7 in \cite{Numerical-Methods})
throughout most of the domain of integration. This gauge is defined
by

\begin{equation}
\sigma_{max}(u)=0\,,\,\:\sigma_{v}(v)=0\,,\label{eq: Sigma_max:u+v}
\end{equation}

\noindent where $\sigma_{v}(v)\equiv\sigma(u_{0},v)$, namely the
initial value of $\sigma$ along the outgoing characteristic ray;
and $\sigma_{max}(u)$ is defined as the maximal value of $\sigma$
(in the range $v_{0}\leq v\leq v_{max}$) along each ray of constant
$u$. \footnote{The gauge condition $\sigma_{max}(u)=0$ translates into a choice
of concrete initial value $\sigma(u,v_{0})$ through an extrapolation
procedure explained in subsections 7.1 of \cite{Numerical-Methods}.
The same type of interpolation procedure also applies to the ``singularity
approach'' variant described below.} This entirely resolves the aforementioned numerical problem at the
event-horizon (as well as an analogous problem which may arise at
the inner horizon), as demonstrated in Ref. \cite{Numerical-Methods}.

\subsection{The ``singularity approach'' gauge variant\label{sub:The-singularity-approach}}

\noindent At late $u$ values, somewhere beyond the EH, we switch
to a new gauge condition (for $u$), for a reason which we now explain: 

As was mentioned in the Introduction, in the presence of self-gravitating
scalar field the energy flux across the CH causes the latter to contract:
$r$ steadily decreases from $r_{-}$ down to full contraction at
$r=0$, where the shrinking CH intersects the $r=0$ spacelike singularity
(see Fig. \ref{fig:NL-num-grid}). One of the interesting issues which
has not been addressed yet is the asymptotic behavior of the shrinking
CH (and its neighborhood) on approaching the point of full contraction.
This topic is in principle amenable to numerical investigation, provided
that the numerical code would be capable of effectively resolving
the shrinking CH close to full contraction. 

A straightforward approach for improving numerical resolution near
full contraction would be to refine the steps in $u$ while $r$ shrinks
along the CH. Our basic strategy, however, is to achieve such an effective
refinement by controlling the gauge condition for $u$ --- as we did
earlier for successfully crossing the EH and IH. To this end, we developed
a special variant of $u$-gauge, the \emph{singularity-approach gauge}.
This gauge coincides with the maximal-$\sigma$ gauge up to a certain
$u$ value inside the BH, which we denote $u_{s}$. Beyond $u=u_{s}$
we further decrease $\sigma$ (which effectively amounts to refinement
of the $u$ coordinate), by an amount that depends on the smallness
of $r$ at the CH. We find that a useful choice for the new $\sigma$
is via $e^{\sigma_{max}(u)}\propto r(u,v_{max})^{2}$. (We point out
that in our numerical simulations $v_{max}$ is sufficiently large,
such that it effectively represents the CH itself.) The gauge condition
at $u>u_{s}$ is thus 
\begin{equation}
\sigma_{max}(u)=2\ln\left[r(u,v_{max})/M_{0}\right]+const\,\,,\,\,\:\sigma_{v}(v)=0\,.\label{eq:gauge-fix}
\end{equation}
The $const$ is fixed by requiring continuity of $\sigma_{max}$ at
$u=u_{s}$, namely $-2\ln\left[r(u_{s},v_{max})/M_{0}\right]$. 

We point out that for the main objective of this paper --- exploring
the shock structure along the outgoing IH --- the fine resolution
of the CH close to full contraction is not needed. We could actually
use the original maximal-$\sigma$ gauge (\ref{eq: Sigma_max:u+v})
throughout the domain of integration. However, we designed our numerical
code as a multi-purpose tool, and for this reason we chose to implement
the singularity-approach gauge variant in the code.

\subsection{Presentation of numerical results}

As was mentioned above, in our numerical simulations we used several
grid refinement levels, $N=80,160,320,640$, to test convergent rate
and final accuracy. Throughout the paper, in all figures which display
data along individual timelike geodesics, we plot the data for both
$N=640$ (solid curves) and $N=320$ (dashed curves). In all figures
but one, there is an excellent agreement between these two resolutions
and the two curves effectively overlap, hence the dashed curves cannot
be noticed. The only exception is the highly zoomed figure \ref{fig:r-shock-zoom},
where the dashed lines can barely be noticed (but still, the deviations
are small). In figures which display processed timelike geodesics
data (such as \ref{fig:test-phi-shock-wave-developement-2}, \ref{fig:nonlinear-phi-shock-wave-developement-2}
or \ref{fig:r-width-params}) or data along null geodesics (figures
\ref{fig:3d-Null-Geodesic-Formation} and \ref{fig:3d-RN-comparison}),
we plot the data from the best resolution ($N=640$) only.

Note also that in all graphs below, we use units in which the initial
RN mass parameter is $M_{0}=1$ (in addition to $c=G=1$), and we
also set $u_{0}=v_{0}=0$.

\section{TEST SCALAR FIELD \label{sec:test-scalar-field}}

In this section we consider the evolution of a test scalar field on
a prescribed RN background, and numerically explore the evolving shock-wave
in this field. For convenience we denote this test field by $\phi$,
to distinguish it from its self-gravitating counterpart $\Phi$. This
field satisfies the same field equation as $\Phi$, namely $\Square\phi=0$
{[}and hence same equation as (\ref{eq:  phi_evolutio}){]}. However,
it has no contribution to the Einstein equations, hence the geometry
remains RN throughout. 

We first describe the set-up of initial data for $\phi$. Then we
introduce a family of timelike infalling geodesics, which probe $\phi$
inside the BH as a function of their proper time $\tau$ and thereby
expose the effective shock-wave structure. Then we move on to analyze
the shock properties, and in particular how its width decreases with
infall time.

\subsection{Basic parameters and initial conditions\label{sub:Test-Initial-Conditions}}

The test scalar field propagates in the RN geometry, which in Schwarzschild
coordinates reads

\begin{equation}
ds^{2}=-f(r)dt^{2}+f(r)^{-1}dr^{2}+r^{2}d\Omega^{2}\,,\label{eq:metric}
\end{equation}
where $f\equiv1-2M/r+Q^{2}/r^{2}$. The EH and IH are located at 
\begin{equation}
r_{\pm}=M\pm\sqrt{M^{2}-Q^{2}}\,.\label{eq:r_plus_minus}
\end{equation}
The surface gravity at the IH and EH is given by
\[
\kappa_{\pm}=\frac{\sqrt{M^{2}-Q^{2}}}{r_{\pm}^{2}}\,.
\]
We choose here mass parameter $M=1$ \footnote{As was already mentioned in Sec. \ref{sec:physical-system}, our code
actually solves the electrovac Einstein equations and numerically
construct the RN metric (in the chosen gauge). Consequently there
is a small numerical drift in $M$. This drift, however, is of order
$\sim2\cdot10^{-7}$ and does not affect any of our results.} (which actually reflects our choice of units), and charge $Q=0.92$.\footnote{This value of $Q$ was chosen to obtain $\kappa_{-}$ value close
to $1$.} The event and inner horizon values are then $r_{+}\simeq1.392$ and
$r_{-}\simeq0.608$ respectively, and the corresponding IH surface
gravity is $\kappa_{-}\simeq1.06$. 

The location of the numerical grid with respect to the RN background
is illustrated in Fig. \ref{fig:RN-num-grid}. While the outgoing
initial ray $u=u_{0}$ is located outside the EH, the ingoing ray
$v=v_{0}$ penetrates into the BH and subsequently crosses the IH
as well. In practice we set the values of the two initial null rays
to be $u_{0}=v_{0}=0$ for convenience. 

The initial value for the metric variable $\sigma$ was chosen so
as to conform with the ``singularity approach'' gauge condition
described in Sec. \ref{sub:The-singularity-approach}. In particular,
$\sigma$ vanishes along the $u=u_{0}$ initial ray. The initial data
for the scalar field are taken to be those of ingoing pulse at $u=u_{0}$,
as schematically shown in Fig. \ref{fig:RN-num-grid} (in particular
$\phi$ vanishes along $v=v_{0}$). The ingoing pulse on $u=u_{0}$
has a finite support at $v_{1}\leq v\leq v_{2}$, and we choose it
such that both $\phi$ and its derivative are smooth at the edges
$v_{1,2}$. Specifically we take the symmetric initial pulse 

\begin{center}
\begin{equation}
\phi(u_{0},v)=\phi_{0}(v)\equiv\begin{cases}
\begin{array}{c}
\frac{64(v-v_{1})^{3}(v_{2}-v)^{3}}{(v_{2}-v_{1})^{6}}\\
0
\end{array} & \begin{array}{c}
|\,v_{1}\leq v\leq v_{2}\\
|\,\,\,\,\,otherwise
\end{array}\end{cases}\label{eq:SF_V}
\end{equation}

\par\end{center}

\noindent The polynomial form was selected due to its simplicity.
The pre-factor $64$ was introduced such that the ingoing pulse has
a unit amplitude. \footnote{As was already noted in footnote \ref{fn: test-field}, in the numerical
run we actually multiply this initial function by an extremely small
overall factor $A_{0}=10^{-20}$ (so as to numerically ``mimic''
a test field). Throughout this section all the results for $\phi$
are presented with this artificial pre-factor scaled-out (namely,
we divide the actual numerical data by $A_{0}$).} In our test-field runs we chose $v_{1}=1,\,v_{2}=3$. Overall the
domain of integration was taken to be $u_{0}=v_{0}=0$, $u_{max}=105$,
and $v_{max}=120$. The value of $r$ at the initial vertex was $r(u_{0},v_{0})=5$,
and along the outgoing initial ray it grew monotonically up to $r(u_{0},v_{max})\simeq52.8$. 

The numerical code was then run with the four resolutions $N=80,160,320,640$,
in order to control accuracy and convergence rate, as described in
Sec. \ref{sec:Basic-Numerical-Algorithm}.

\subsubsection{Double-null Eddington coordinates}

We denote the Double-null Eddington coordinates by $u_{e}$ and $v_{e}$.
In the RN background (outside the BH) they are defined by 
\[
v_{e}\equiv t+r_{*}\,,\,\,u_{e}\equiv t-r_{*}\,,
\]
where $r*$ is the tortoise coordinate given by
\begin{equation}
r_{*}(r)=r+\frac{r_{+}^{2}}{r_{+}-r_{-}}\ln|r-r_{+}|-\frac{r_{-}^{2}}{r_{+}-r_{-}}\ln|r-r_{-}|+const\,.\label{eq:RN_r_star}
\end{equation}
The integration constant in this equation may be chosen at will. Owing
to time-translation symmetry these coordinates are defined up to a
global displacement. We find it convenient to set $u_{e}=0$ at the
initial ray $u=u_{0}$, and to set $v_{e}=0$ at the end of the injected
pulse, namely at $v=v_{2}$. This choice fixes the integration constant
in Eq. (\ref{eq:RN_r_star}), by requiring $r_{*}=0$ at $r(u_{0},v_{2})$.

We shall hereafter reserve the symbols $u,v$ to the double-null coordinates
that we actually use in the numerical simulation (namely, those defined
by the gauge conditions prescribed in Sec. \ref{sub:The-singularity-approach}).
Later we shall need to use the relation $v_{e}(v)$. This function
is easily determined numerically, at the outgoing initial ray $u=u_{0}$,
using $v_{e}-u_{e}=2r_{*}$. Since $u_{e}$ vanishes at that ray by
definition, we obtain
\begin{equation}
v_{e}(v)=2r_{*}\left(r(u_{0},v)\right)\,.\label{eq:ve(v)}
\end{equation}

\subsection{Family of timelike geodesics \label{sub:Teste-Radial-Geodesics}}

Following Ref. \cite{Marolf-Ori_Shockwave}, we consider here a family
of timelike geodesics related to each other by time translation, and
use them to probe the BH interior. We shall refer to such a family
as a \emph{time-translated set of geodesics} (TTSG). In the present
case of test perturbation, this family consists of exact RN geodesics,
all sharing the same energy $(E=-u_{t})$ and angular momentum $(L=u_{\varphi})$
values, which fall into the BH. Each geodesic is marked by a typical
timing parameter. We adopt here the same timing parameter as in \cite{Marolf-Ori_Shockwave},
namely $v_{eh}$ --- the value of Eddington null coordinate $v_{e}$
at the moment of EH crossing. 

For concreteness and simplicity, we choose here the set of geodesics
$E=1,\,L=0$, representing a family of infalling observers on marginally-bound
radial geodesics. These observers will probe the scalar field inside
the BH as a function of their proper time $\tau$ and infall time
parameter $v_{eh}$. We set the proper time value $\tau=0$, at each
geodesic, at the moment of EH crossing. The parameter $v_{eh}$ for
each geodesic is determined from the value of $v$ at horizon crossing,
via Eq. (\ref{eq:ve(v)}).

Our numerical code produces values of the unknowns $r,\sigma,\Phi$
on a discrete set of grid points in the ($u,v$) coordinates, as described
in Sec. \ref{sec:Basic-Numerical-Algorithm}. However, for the shock-wave
analysis we shall have to evaluate these functions on the aforementioned
radial timelike geodesics. The latter may be considered as a set of
bent curves $v(u)$, obtained from the geodesic equation (see Appendix
\ref{sub:Radial-Timelike} for more details). Generically, these curves
do not pass at grid points. Therefore, at each grid value of $u$
we interpolate our numerical results (given on the discrete $v$ values)
to the desired value $v(u)$ on the geodesic. To this end we use second-order
interpolation.

\subsection{Shock wave in the scalar field \label{sub:Test-Shock-Wave-in-Phi}}

The effective shock-wave formation in the test scalar field $\phi$
is demonstrated in Fig. \ref{fig:test-phi-shock-wave-formation},
which displays $\phi(\tau)$ for various infalling geodesics, labeled
by their $v_{eh}$ values. In these geodesics, whereas $r$ monotonically
decreases towards $r_{-}$, the scalar field $\phi$ is non-monotonic:
it decreases at some stage but then increases again (this is better
seen in panel b, in which $\tau$ is shifted to enable visual separation
of the different geodesics). The evolving shock structure manifests
itself as a rapid sharpening (with increasing $v_{eh}$) of both the
decrease phase and the subsequent increase phase. The increase phases
in all geodesics (except the earliest ones, say $v_{eh}=4-5$) form
an effectively vertical line. These vertical lines, in all geodesics,
occur approximately at the same proper time ($\tau\simeq0.7438$),
hence they all overlap in panel a, forming the ``wall like'' structure.
Also, these vertical sections occur in all geodesics at approximately
the same $r$ value, namely $r=r_{-}$. \footnote{Note that in the way we set the family of geodesics, they all share
the same function $r(\tau)$, due to time-translation invariance. } 

In order to enable visual separation of the different geodesics, such
that the shock (and pre-shock) phase of the individual geodesics would
be seen, in panel b the $\tau$ value of each geodesic is shifted
by a certain amount, proportional to its $v_{eh}$.

Whereas Fig. \ref{fig:test-phi-shock-wave-formation} demonstrated
the shock formation phase by presenting relatively early geodesics
($v_{eh}\le15$), Fig. \ref{fig:test-phi-shock-wave-developement-1}a
displays much later infalling geodesics $20\le v_{eh}\le100$. The
shock structure is now fully developed: It takes the form of a vertical
line, located at the same proper time for all geodesics. It is placed
just at the $\tau$ value that corresponds to $r=r_{-}$ (the dashed
vertical line). 

\begin{center}
\begin{figure}[H]
\begin{centering}
\includegraphics[scale=0.25]{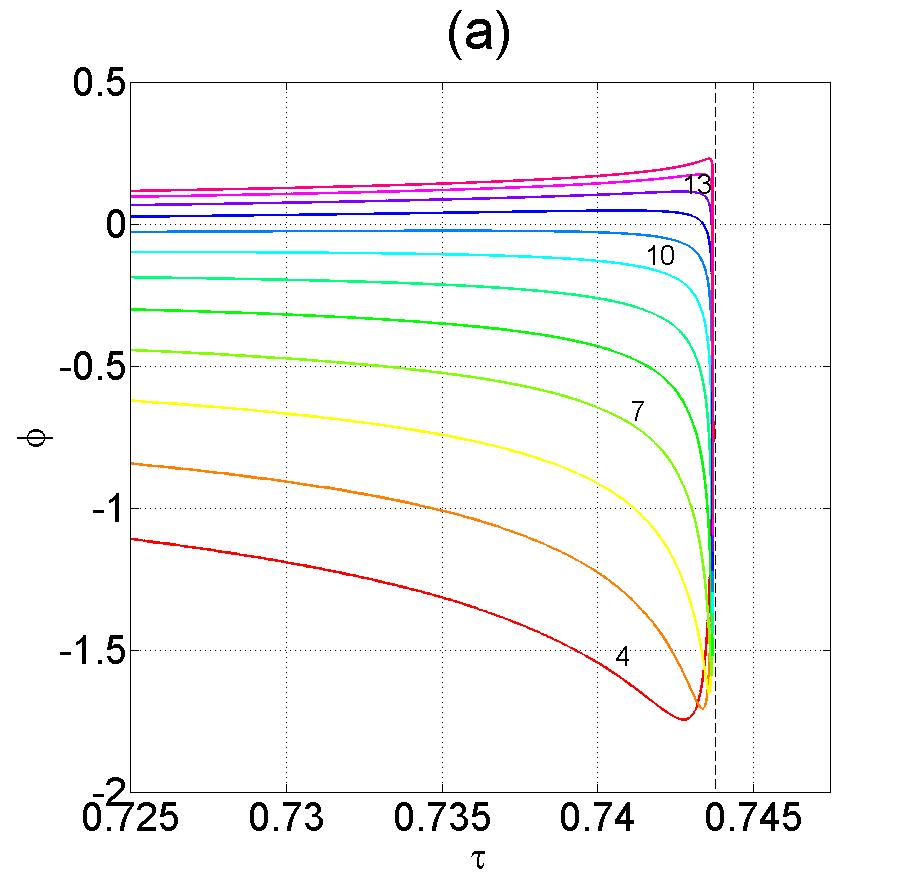}\includegraphics[scale=0.25]{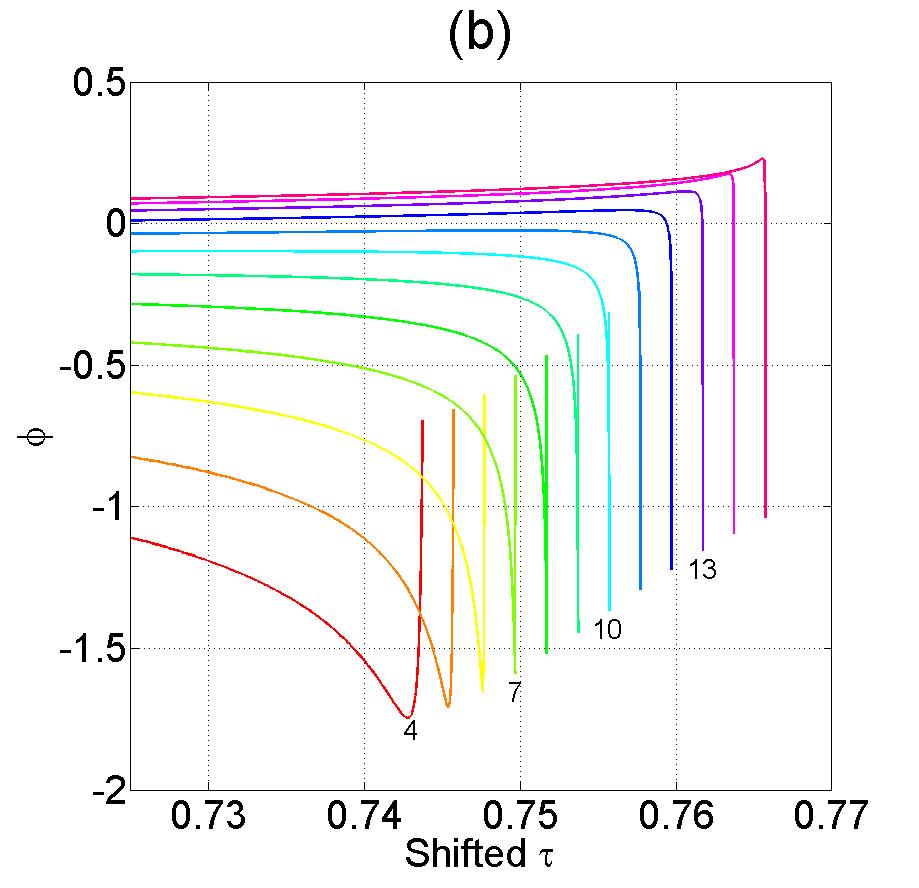}
\par\end{centering}

\protect\caption{Shock-wave formation in the test scalar field $\phi$. Both panels
present $\phi$ as a function of proper time $\tau$ along a family
of radial timelike geodesics with $E=1$. Panel a uses true $\tau$
values at each geodesic (with $\tau=0$ set at EH crossing); panel
b uses $\tau$ values artificially shifted by an amount $(i-1)\times2\cdot10^{-3}$
for the $i$-th geodesic ($i=1,...,12$, corresponding to $v_{eh}=4,...,15$),
in order to enable clear visual separation of the different geodesics.
Geodesics are distinguished by different colors and different $v_{eh}$
values. $v_{eh}$ increases by increments of $1$ from the bottom/left
$(v_{eh}=4)$ to the top/right $(v_{eh}=15)$. The vertical dashed
line in panel a represents the RN theoretical $\tau$ value of IH
crossing, $\tau\simeq0.7438$. \label{fig:test-phi-shock-wave-formation}}
\end{figure}

\par\end{center}

\subsubsection{Shock resolution and shock Sharpening\label{sub:Shock-developement1}}

Figures \ref{fig:test-phi-shock-wave-formation}a and \ref{fig:test-phi-shock-wave-developement-1}a
made it clear that the shock structure involves extremely short $\tau$
intervals --- which further shorten with increasing $v_{eh}.$ Still,
we can resolve the shock by using a logarithmic scale for $\tau$.
To this end, however, we shall have to use the shifted variable $|\tau-\tau_{final}|$,
where $\tau_{final}$ is the last proper time value for each geodesic
(achieved at $u=u_{max}$ arrival). Figure \ref{fig:test-phi-shock-wave-developement-1}b
provides such a logarithmic representation of $\phi$ as a function
of $|\tau-\tau_{final}|$, for the same set of geodesics as in Fig.
\ref{fig:test-phi-shock-wave-developement-1}a (namely, $20\le v_{eh}\le100$). 

This figure demonstrates that the shock structure {[}which actually
enfolds alternating domains of decreasing and increasing $\phi(\tau)${]}
is essentially unchanged while $v_{eh}$ increases. The main effect
of the increase in $v_{eh}$ is to shorten the proper-time width of
the configuration. This sharpening is expressed in Fig. \ref{fig:test-phi-shock-wave-developement-1}b
by the overall rightward shift in the logarithmic scale, with increasing
$v_{eh}$. 

\begin{center}
\begin{figure}[H]
\begin{centering}
\includegraphics[scale=0.25]{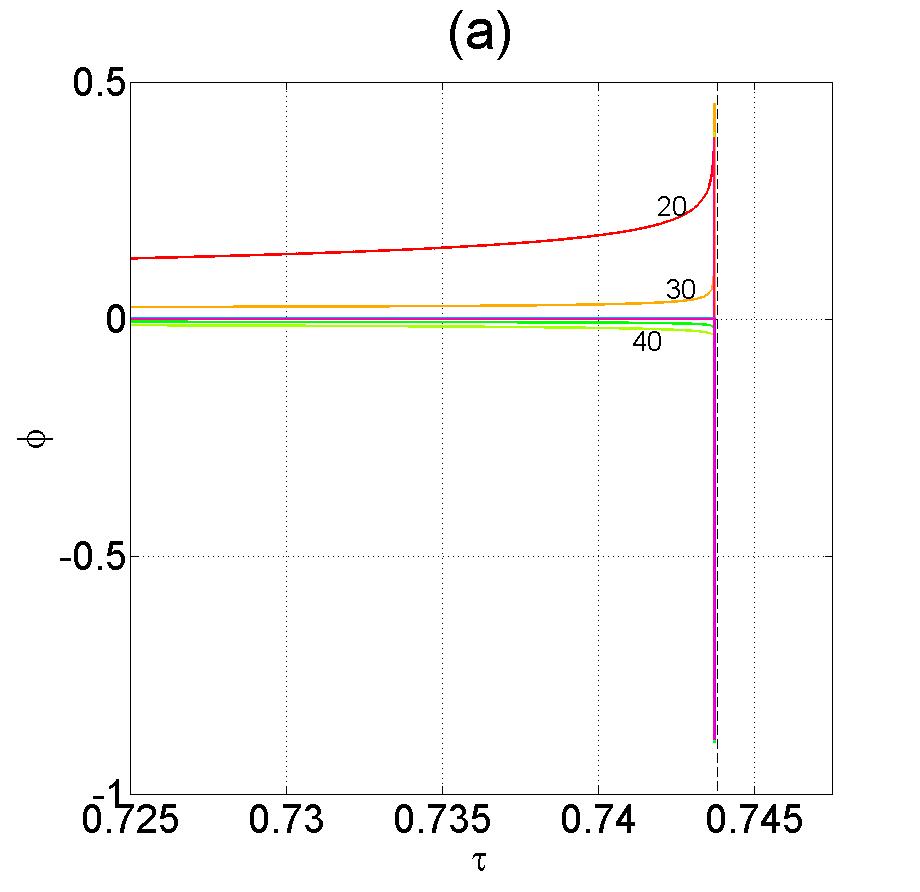}\includegraphics[scale=0.25]{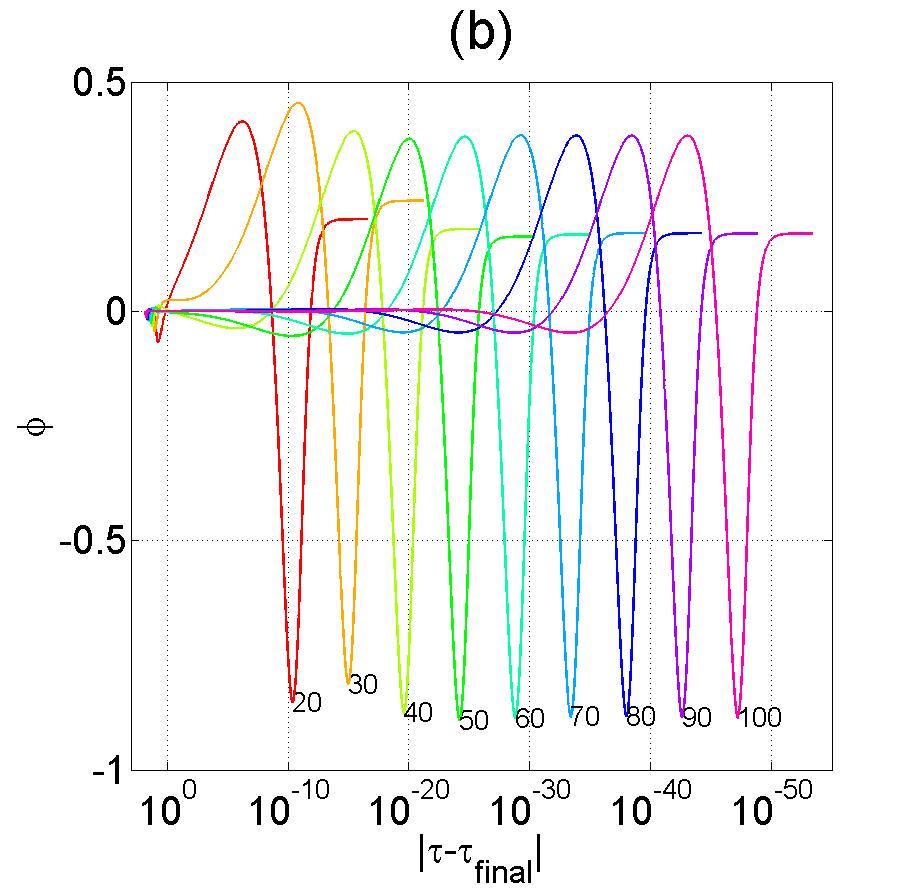}
\par\end{centering}

\protect\caption{Shock-wave structure at late time. The graph displays $\phi(\tau)$
along a family of radial timelike geodesics with $E=1$, this time
for later geodesics, $20\le v_{eh}\le100$. The different geodesics
are again marked by different colors and different $v_{eh}$ values;
$v_{eh}$ increases by increments of $10$ from top/left $(v_{eh}=20)$
to bottom/right $(v_{eh}=100)$. Panel a uses a linear scale in $\tau$,
clearly showing the well-developed shock structure (the vertical line
at the right); panel b uses a logarithmic scale in $|\tau-\tau_{final}|$,
where $\tau_{final}$ is the final $\tau$ value at $u=u_{max}$.
The vertical dashed line in panel a again represents the theoretical
$\tau$ value of IH crossing in RN, $\tau\simeq0.7438$. Panel b exhibits
the sharpening of the shock --- the width of the scalar field pulse
exponentially decreases with $v_{eh}$, as inferred from the apparently
``rigid'' shift to the right (with respect to the logarithmic scale)
with increasing $v_{eh}$. \label{fig:test-phi-shock-wave-developement-1}}
\end{figure}

\par\end{center}

According to the analysis of Ref. \cite{Marolf-Ori_Shockwave}, the
decrease in any proper-time width scale $\Delta\tau$ associated with
the scalar-field shock-like signal should be exponential (for sufficiently
late geodesics): $\Delta\tau\propto e^{-\kappa v_{eh}}$, where $\kappa\equiv\kappa_{-}$
is the surface gravity at the inner horizon. To test this prediction
we chose here a specific width scale $\Delta\tau$: the difference
in $\tau$ between the minimum and maximum points in $\phi(\tau)$
\footnote{To be more specific, these are the (last) maximum and minimum points
in the profile shown in Fig. \ref{fig:test-phi-shock-wave-developement-1}b
for each geodesic. } at each geodesic. The results are presented in Fig. \ref{fig:test-phi-shock-wave-developement-2}.
As expected $\Delta\tau$ decays exponentially, in nice agreement
with the theoretically predicted rate $e^{-\kappa v_{eh}}$ \footnote{Similar results were obtained for another choice of typical width,
the \emph{full width at half maximum} of $\phi(\tau)$ at the various
geodesics. }.

\begin{center}
\begin{figure}[H]
\begin{centering}
\includegraphics[scale=0.25]{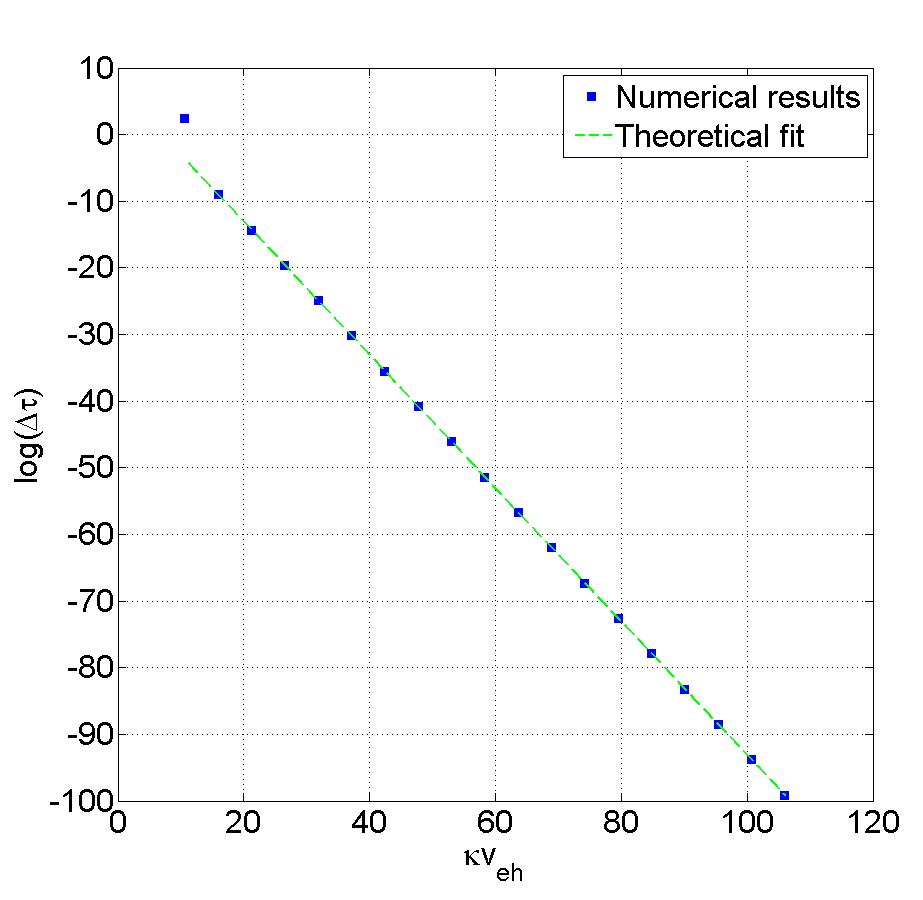}
\par\end{centering}

\protect\caption{Exponential sharpening rate of the test scalar field shock, as demonstrated
by the decrease of $\Delta\tau$ with increasing $v_{eh}$. Each point
represents a single $E=1$ radial timelike geodesic; the geodesics
are in the range $10\leq v_{eh}\leq100$, with increments of $5$.
The straight diagonal dashed green line is the theoretical fit, $\ln(\Delta\tau)=-\kappa v_{eh}+const$.
All geodesics show excellent agreement with the theoretical fit, except
the first geodesic $v_{eh}=10$ (which is seemingly too early for
the late-time theoretical relation to hold). \label{fig:test-phi-shock-wave-developement-2}}
\end{figure}

\par\end{center}

Note that Marolf and Ori mainly focused in their shock-wave analysis
\cite{Marolf-Ori_Shockwave} on the late time domain, where the background
geometry is approximately static. However, as Fig. \ref{fig:test-phi-shock-wave-formation}
demonstrates, the shock-like behavior can also be detected for very
early infalling geodesics, even for e.g. $v_{eh}=4$. \footnote{We point out that the observation of the shock-like behavior is obviously
scale dependent: clearly, a sufficient zoom on the $\tau$ scale in
e.g. Fig. \ref{fig:test-phi-shock-wave-formation} would resolve the
sharp features. We argue, however, that since the $\tau$ scale in
Fig. \ref{fig:test-phi-shock-wave-formation} is already rather small
(the overall horizontal scale of panel a is $\approx0.025$, which
is $\ll1$), the sharp features seen in that panel indeed indicate
a shock-like behavior.} Notice, in addition, that the nice agreement of the points in Fig.
\ref{fig:test-phi-shock-wave-developement-2} with the straight diagonal
dashed line (representing the theoretical sharpening rate $\Delta\tau\propto e^{-\kappa v_{eh}}$)
starts already at $v_{eh}\approx15$.

\section{SELF-GRAVITATING SCALAR FIELD PERTURBATION \label{sec:self-gravitating-scalar-field}}

In the case of self-gravitating scalar perturbation, the shock wave
manifests itself in the scalar field $\Phi$ and also in the metric.
In this section, after describing the set-up of the problem, we shall
first present the shock in $\Phi$, and subsequently the shock in
the metric function $r$. Then we shall re-analyze the shock in $r$
using a family of ingoing null geodesics (rather than timelike ones).

\subsection{Set-up and initial conditions\label{sub:Nonlinear-Initial-Conditions}}

The setup in the present case is similar to the test-field case presented
in Sec. \ref{sec:test-scalar-field}, except that our scalar field
is now self-gravitating. Initially (at $v<v_{1})$ we have an exact
RN geometry with mass $M_{0}=1$ and charge $Q=0.95$. Then at $v_{1}<v<v_{2}$
we inject the ingoing scalar-field pulse, as schematically shown in
Fig. \ref{fig:NL-num-grid}. (Note that $\Phi$ vanishes at the ingoing
initial ray $v=v_{0}$.) The scalar field then spreads throughout
the domain $v>v_{1}$. At later times the field back-scatters and
decays, and eventually we have an approximate RN geometry at large
$v$, with the same charge $Q=0.95$ but with final mass $m_{final}>M_{0}$. 

The injected scalar field pulse is taken to be of the form 
\[
\Phi(u_{0},v)=A\,\phi_{0}(v)\,,
\]
where $\phi_{0}(v)$ is the basic finite-support pulse function given
in Eq. (\ref{eq:SF_V}), and $A$ is a free amplitude parameter. We
chose here the values $A=0.115$ along with pulse boundary parameters
$v_{1}=1,\,v_{2}=7$. With this choice the final mass becomes $m_{final}\cong1.4587$.
\footnote{In principle $m_{final}$ should be defined as the limit of the mass
function $m$ along the EH as $v\to\infty$. It practice we only monitor
the evolution up to $v_{max}$, hence we take $m_{final}$ to be the
mass function evaluated at the EH at $v=v_{max}$. We verified, by
inspecting the large-$v$ freezing of $m(v)$ along the horizon, that
the change in this function at $v>v_{max}$ is negligible, probably
smaller than $10^{-6}$.\label{fn:m_final_drift}} The corresponding inner-horizon surface gravity parameter is $\kappa\simeq8.94$.
We picked these parameters (despite the relatively large resultant
value of $\kappa$) in order to achieve significant focusing of the
CH within the numerical domain. \footnote{The focusing effect, in this context, is a gradual decrease of $r$
along the CH. While we do not actually reach the CH in our numerical
simulation (it corresponds to $v\to\infty$), we get relatively ``close''
to it in terms of the behavior of the various fields, and in particular
$\Phi$ and $r$. With the choice $A=0.115$ we get a significant
focusing in our numerical domain: $r$ shrinks along $v=v_{max}$
down to $\approx10^{-2}$ or even less. This significant focusing
is advantageous as it allows a better visibility of the shock behavior
in $r$ (discussed in Sec. \ref{subsec:shock-in-r}). }

The initial conditions for the metric function $\sigma$ are set in
accord with the ``singularity approach'' gauge condition presented
in Sec. \ref{sub:The-singularity-approach}. In particular, $\sigma$
vanishes along the $u=u_{0}$ initial ray. The initial data for $r$
are calculated, along both initial rays, from the constraint equations.
As before, we set $u_{0}=v_{0}=0$ for convenience. The other boundaries
of the domain of integration are $u_{max}=182$ and $v_{max}=120$.
The value of $r$ at the initial vertex is again $r(u_{0},v_{0})=5$,
and along the outgoing initial ray it grows monotonically up to $r(u_{0},v_{max})\simeq44.0$.

Note that there are two different RN phases in spacetime: (i) an \emph{exact}
initial RN geometry at $v<v_{1}$, and (ii) an \emph{approximate}
RN geometry at late time, namely $v\gg1$.\footnote{A more precise condition would be $v-v_{2}\gg m_{final}$ (but in
the present case the two conditions are essentially the same).} The effective shock wave is essentially a phenomenon that takes place
at late times, hence it is the approximate RN domain (ii) that will
be mostly relevant to the shock analysis below. The event and inner
horizon values of this domain are $r_{+}\simeq2.566$ and $r_{-}\simeq0.352$
respectively.

\begin{center}
\begin{figure}[H]
\begin{centering}
\includegraphics[scale=0.6]{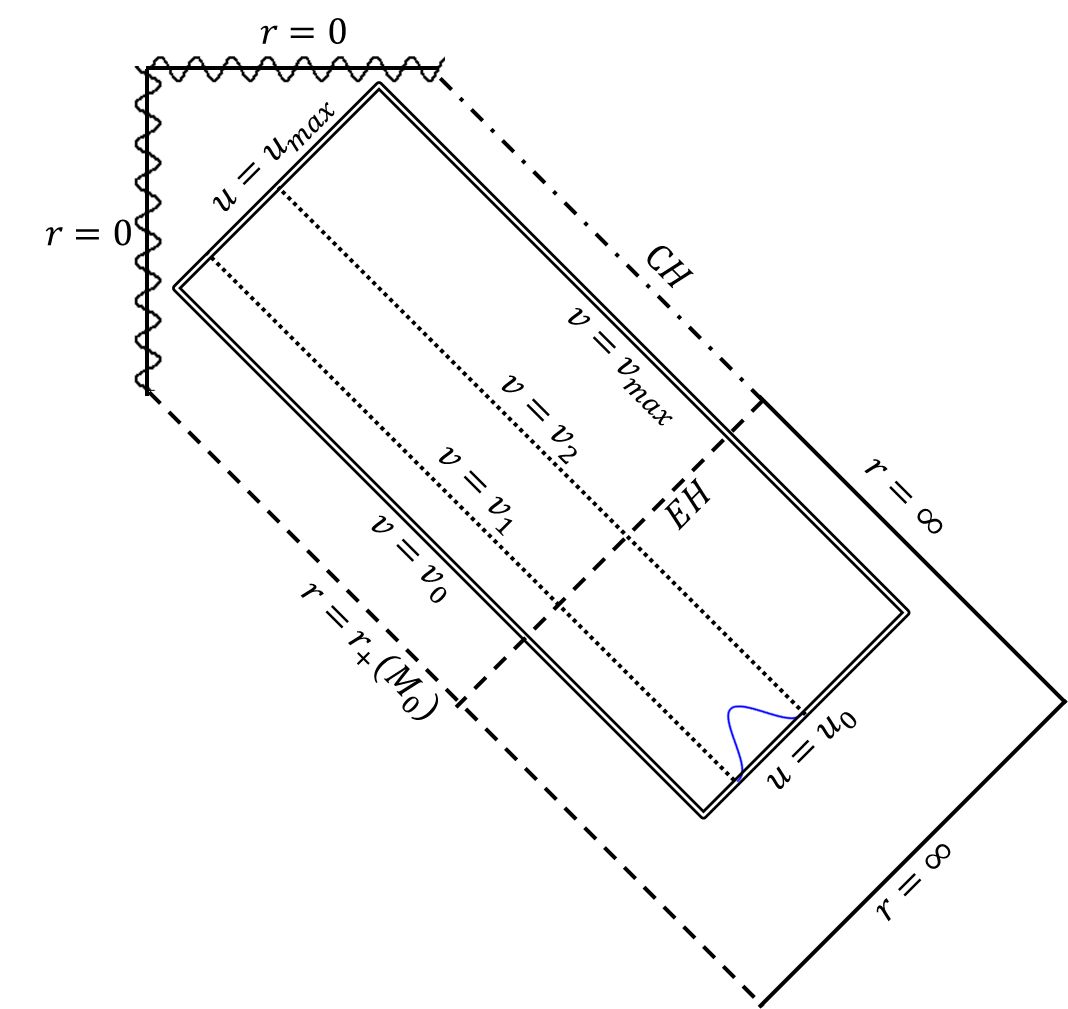}
\par\end{centering}

\protect\caption{\label{fig:NL-num-grid}This figure describes the location of the
numerical domain of integration with respect to the perturbed charged
BH spacetime. Double lines represent the boundary of the numerical
domain $(u=u_{0},\,u=u_{max},\,v=v_{0}$ and $v=v_{max}$); Solid
lines represent future and past null infinity $(r=\infty)$. The left-going
dashed line at bottom/left represents the external horizon ($r=r_{+}(M_{0})$)
of the initial RN geometry at $v<v_{1}$. The right-going dashed line
represents the event horizon of the perturbed BH. The ingoing scalar-field
pulse is drawn schematically on the outgoing initial ray $u=u_{0}$.
The dashed-dotted line at top/right represents the CH inside the BH.
It steadily shrinks due to the scalar-field perturbation, from $r=r_{-}(m_{final})$
down to $r=0$. The vertical wavy line represents the timelike $r=0$
singularity of the initial RN geometry. The horizontal wavy line represents
the spacelike $r=0$ singularity in the perturbed geometry. Note,
however, that these two singularities are located outside the domain
of integration.}
\end{figure}

\par\end{center}

\subsubsection{Family of timelike geodesics \label{sub:SG-Radial-Geodesics}}

In order to probe the shock-like behavior of the various fields, in
the test-field case we employed a family of radial timelike RN geodesics
related to each other by time-translation symmetry, namely radial\emph{
time-translated set of geodesics} (TTSG). Unfortunately, in the self-gravitating
case the geometry is no longer static, and an exact TTSG does not
exist. Nevertheless, since at large $v$ the geometry becomes asymptotically
RN (hence asymptotically static), we can still construct a family
of geodesics which are related by an \emph{approximate} time translation
--- an approximation that improves with increasing $v$. Note that
this construction is not unique, there are many ways to construct
such an approximate TTSG. All we require is that this family will
asymptotically approach a TTSG at large $v$. 

To define such a family of radial timelike geodesics, it will be sufficient
to prescribe the $\dot{r}$ value of each geodesic as it crosses the
initial outgoing ray $u=u_{0}$. We would like to mimic as much as
possible the behavior of radial TTSG in exact RN. In the latter case
we have 
\[
\dot{r}=-\sqrt{E^{2}-(1-\frac{2M}{r}+\frac{Q^{2}}{r^{2}})}\,\,,
\]
which for $E=1$ reduces to $-(2M/r-Q^{2}/r^{2})^{1/2}$. In the self-gravitating
case, we replace the constant $M$ by the mass function $m$. Correspondingly
we set the initial condition for the geodesics at $u=u_{0}$ to be

\begin{equation}
\dot{r}(u_{0},v)=-\sqrt{\frac{2m(u_{0},v)}{r(u_{0},v)}-\frac{Q^{2}}{r(u_{0},v)^{2}}}\,\,\,.\label{eq:geodesics}
\end{equation}
At the large-$v$ limit this one-parameter set of geodesics indeed
approaches the $E=1$ radial TTSG in RN, as desired. 

Here again, for all infalling geodesics we set $\tau=0$ at EH crossing,
as we did in the test-field case.

\subsubsection{Outgoing Eddington-like coordinate $v_{e}$ }

In the test-field case we have used the Eddington coordinate $v_{e}$
(evaluated at EH crossing) to parametrize the infalling geodesics.
However, in the self-gravitating case the double-null Eddington coordinates
are not uniquely defined, as the metric is no longer static. We therefore
introduce here the extended notion of \emph{Eddington-like} outgoing
null coordinate, and denote it as before by $v_{e}$. The function
$v_{e}(v)$ is hereby \emph{defined} by Eq. (\ref{eq:ve(v)}), similar
to the test-field case.\footnote{The function $r_{*}(r)$ appearing in this formula is given in Eq.
(\ref{eq:RN_r_star}). It involves the parameters $r_{+}$ and $r_{-}$,
which in the RN case are given by $M\pm\left(M^{2}-Q^{2}\right)^{1/2}$.
In the self-gravitating case, for the sake of defining $v_{e}$ we
replace $M$ by $m(u_{0},v_{max})\cong1.4591$ (which is very close
to $m_{final}$). } As before we set $u_{e}=0$ at $u=u_{0}$ and $v_{e}=0$ at $v=v_{2}$
(end of pulse injection), which in turn fixes the integration constant
in Eq. (\ref{eq:RN_r_star}) by dictating $r_{*}=0$ at $r(u_{0},v_{2})$.

Once the Eddington-like null coordinate $v_{e}$ is defined, the infalling
timelike geodesics are parametrized by their $v_{eh}$, namely the
value of $v_{e}$ at EH crossing.

Note that in the present case $v_{e}$ vanishes at $v=v_{2}=7$, whereas
the domain of integration starts at $v=v_{0}=0$. The range $0<v<7$
thus corresponds to negative $v_{e}$. Accordingly, some of the infalling
geodesics admit \emph{negative} $v_{eh}$ values, as can be seen in
e.g. Fig. \ref{fig:nonlinear-phi-shock-wave-formation}. (But this
negative value of $v_{e}$ or $v_{eh}$ has no special significance.)

\subsection{Shock wave in the scalar field\label{sub:Nonlinear-Shock-Wave-in-Phi}}

The shock wave in the scalar field is demonstrated in Figs. \ref{fig:nonlinear-phi-shock-wave-formation}
and \ref{fig:nonlinear-phi-shock-wave-developement-1}, by displaying
$\Phi(\tau)$ for various infalling geodesics. The early phase of
shock formation is presented in Fig. \ref{fig:nonlinear-phi-shock-wave-formation}:
Whereas early geodesics with say $v_{eh}<-2$ look smooth, geodesics
with larger $v_{eh}$ exhibit a sharp feature. In fact, this sharp
feature involves a rapid decrease followed by an even more rapid increase:
This can be seen in e.g. the geodesic $v_{eh}=-1.8$. The same behavior
also occurs in later geodesics ($-1.6$ to $-1$), although in these
geodesics the decrease-increase phases are too narrow and cannot be
distinguished without additional zoom. 

Figure \ref{fig:nonlinear-phi-shock-wave-developement-1} shows much
later geodesics, with $20\le v_{eh}\le70$. In the linear scale of
panel a, all the geodesics form a seemingly vertical ``wall'', at
$\tau\approx1.789$. This value agrees, to a very good approximation,
with the proper-time interval $\tau_{RN}\cong1.7889$ to move from
the EH to the IH, for an $E=1$ radial geodesic in an RN geometry
with $Q=0.95$ and mass $m_{final}\cong1.4587$. This proper-time
value $\tau_{RN}$ is marked in Fig. \ref{fig:nonlinear-phi-shock-wave-developement-1}a
by a vertical dashed line.

Note that in the early phase shown in Fig. \ref{fig:nonlinear-phi-shock-wave-formation},
unlike its test-field counterpart in Fig. \ref{fig:test-phi-shock-wave-formation}a,
the different geodesics develop their shocks at notably different
$\tau$ values. This directly results from the dynamical character
of the nonlinearly-perturbed spacetime, which breaks time-translation
invariance. On the other hand, in the late-time phase (Fig. \ref{fig:nonlinear-phi-shock-wave-developement-1}a)
the spacetime is approximately static, hence the shocks of all geodesics
fall approximately at the same $\tau$ value, $\thickapprox1.789$. 

\begin{center}
\begin{figure}[H]
\begin{centering}
\includegraphics[scale=0.25]{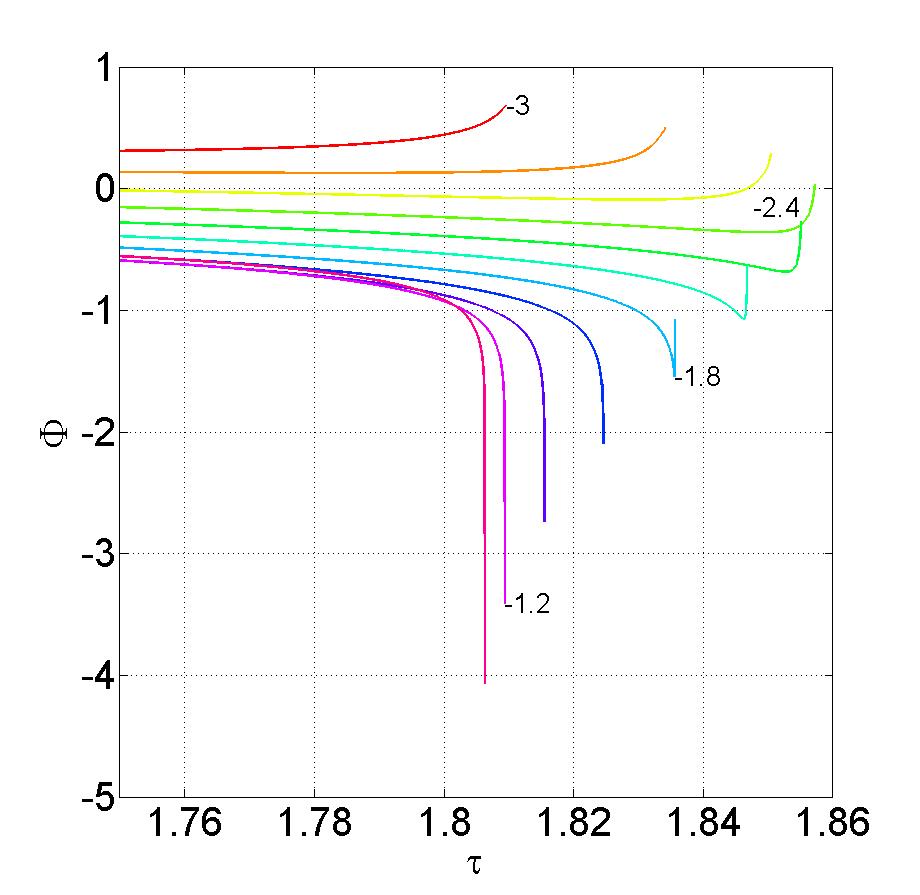}
\par\end{centering}

\protect\caption{Numerical results for the self-gravitating scalar field $\Phi$ as
a function of proper time $\tau$ along the family of radial timelike
geodesics described in the main text. The geodesics are distinguished
by different colors and different values of $v_{eh}$ (the geodesic's
$v_{e}$ value at EH crossing); $v_{eh}$ increases by increments
of $0.2$ from the top of the graph $(v_{eh}=-3)$ to the bottom $(v_{eh}=-1)$.
We set $\tau=0$ at the EH crossing event for each geodesic. The shock
formation process manifests itself as a gradual sharpening of the
scalar field profile with increasing $v_{eh}$. The smaller value
of the increments (compared to the test-field case shown in Fig. \ref{fig:test-phi-shock-wave-formation}),
and the negative values of $v_{eh}$ in the present case, imply a
more rapid shock formation process in the self-gravitating case. This
may be explained by the higher $\kappa$ value in the present case
($8.94$ compared to $1.06$). \label{fig:nonlinear-phi-shock-wave-formation}}
\end{figure}

\par\end{center}

\subsubsection{Shock resolution and shock sharpening \label{sub:Nonlinear-Phi-Shock-Developement}}

To resolve the shock structure, in Fig. \ref{fig:nonlinear-phi-shock-wave-developement-1}b
we use a logarithmic horizontal scale. To this end we again employ
the shifted variable $|\tau-\tau_{final}|$, where, recall, $\tau_{final}$
is the last proper time value for each geodesic (at $u=u_{max}$ arrival).
This panel shows that in the self-gravitating case too, for sufficiently
late geodesics the shock shape is essentially unchanged while $v_{eh}$
increases --- apart from a uniform rightward motion, which expresses
an overall shrinking of the shock's $\tau$ scale. 

To quantify the rate of shock sharpening, we again resort to the width
parameter $\Delta\tau$, defined to be the difference in $\tau$ between
the minimum and maximum points in $\Phi(\tau)$ (those seen in Fig.
\ref{fig:nonlinear-phi-shock-wave-developement-1}b) at each geodesic.
Figure \ref{fig:nonlinear-phi-shock-wave-developement-2} displays
$\Delta\tau$ of the various geodesics as a function of their infall
time $v_{eh}$. It again shows excellent agreement with the exponential
sharpening rate $\Delta\tau\propto e^{-\kappa v_{eh}}$ (represented
by the straight diagonal dashed line), theoretically predicted in
Ref. \cite{Marolf-Ori_Shockwave} (for late-infall geodesics). Exceptional
are the two earliest geodesics shown in the graph, namely $v_{eh}=0$
and $v_{eh}=5$.

\begin{center}
\begin{figure}[H]
\begin{centering}
\includegraphics[scale=0.25]{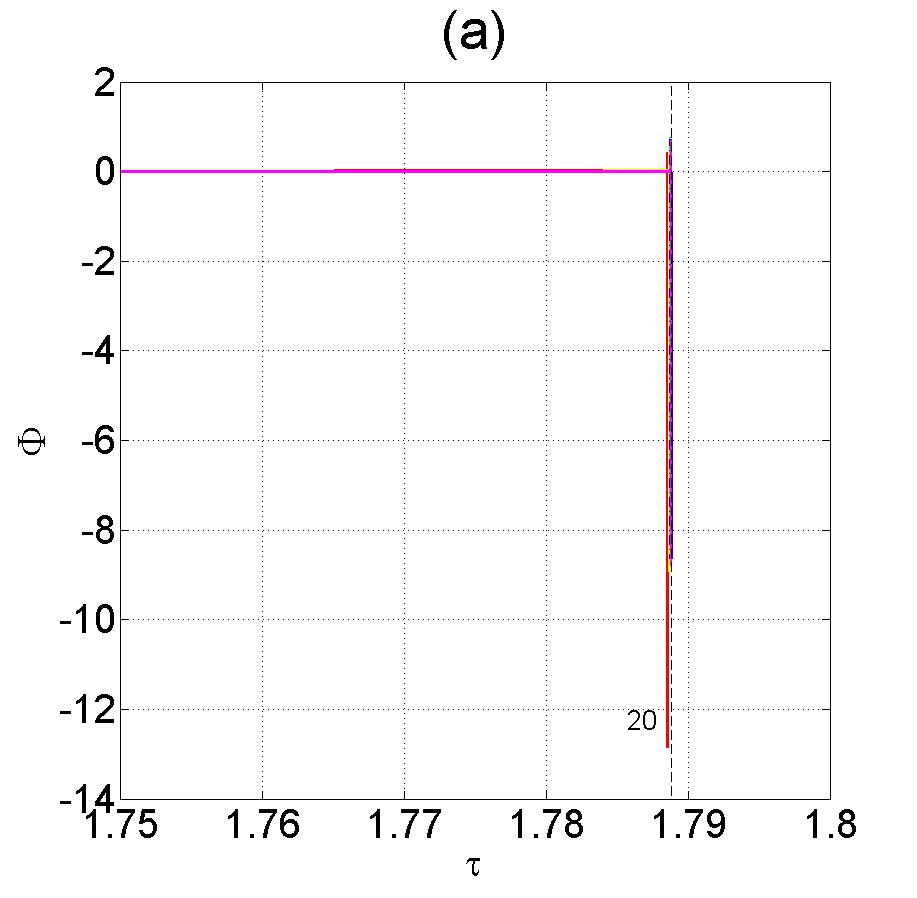}\includegraphics[scale=0.25]{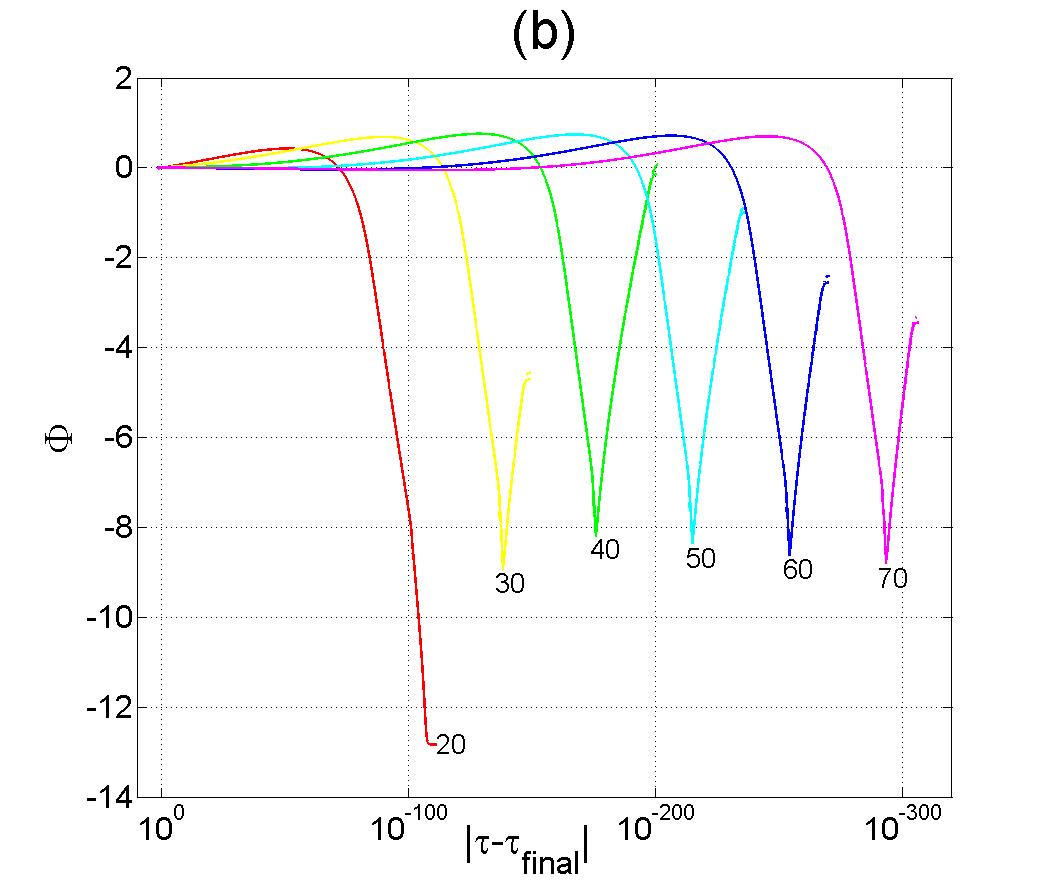}
\par\end{centering}

\protect\caption{Numerical results for $\Phi$ as a function of proper time $\tau$
along a family of radial timelike geodesics. Panel a employs a linear
scale in $\tau$; panel b employs a logarithmic scale in $|\tau-\tau_{final}|$,
to allow shock resolution. Both panels display the same set of geodesics.
The different geodesics are marked by different colors and different
$v_{eh}$ values; $v_{eh}$ increases by increments of $10$ from
bottom/left ($v_{eh}=20$) to top/right ($v_{eh}=70$). The vertical
dashed line in panel a represents $\tau_{RN}\protect\cong1.7889$,
which is the proper-time moment of IH crossing for a marginally-bound
radial geodesic in RN spacetime with $Q=0.95$ and mass $m_{final}=1.4587$.
Panel b exhibits the sharpening of the shock: The width of the scalar-field
profile seems to decrease exponentially with $v_{eh}$. It may also
be seen that the shape of the scalar-field profile at late times (say
$v_{eh}\gtrsim40$) is essentially unchanged. \label{fig:nonlinear-phi-shock-wave-developement-1}}
\end{figure}

\par\end{center}

\begin{center}
\begin{figure}[H]
\begin{centering}
\includegraphics[scale=0.25]{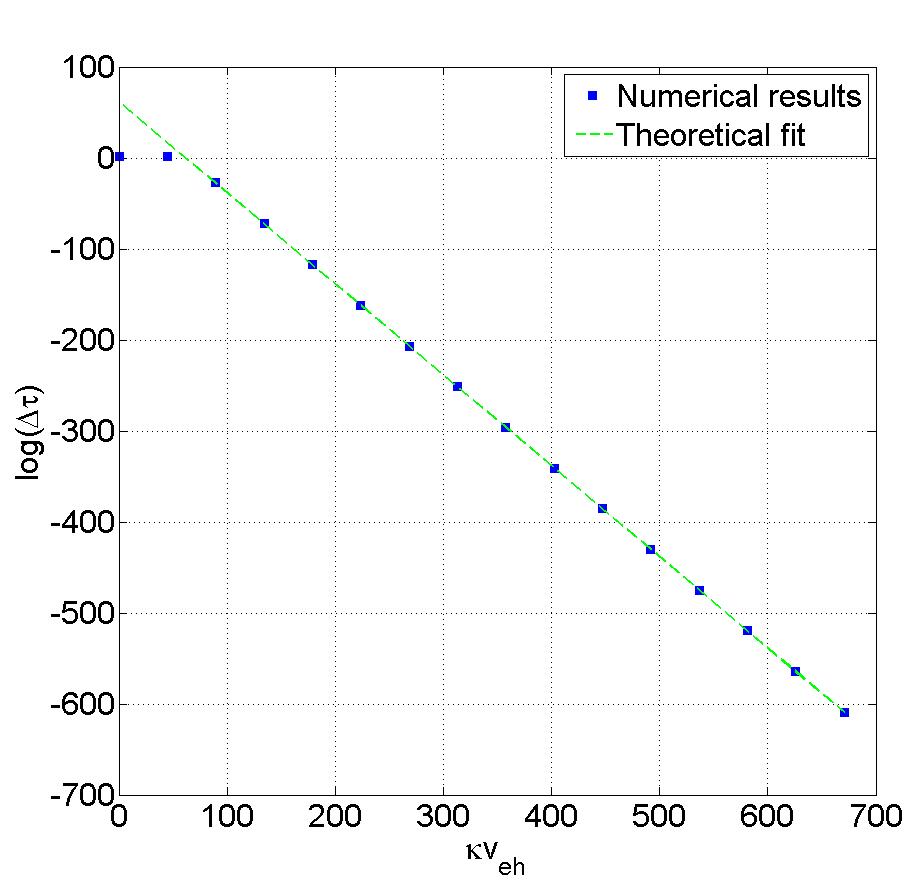}
\par\end{centering}

\protect\caption{\label{fig:nonlinear-phi-shock-wave-developement-2} Exponential sharpening
rate of the self-gravitating scalar field shock, as demonstrated by
the decrease of $\Delta\tau$ with increasing $v_{eh}$. Each point
represents a single timelike radial geodesic; the geodesics are in
the range $0\leq v_{eh}\leq75$, with increments of $5$. The straight
dashed green line is the theoretical fit, $\ln(\Delta\tau)=-\kappa v_{eh}+const$.
All geodesics show excellent agreement with the theoretical fit, except
the two earliest ones ($v_{eh}=0,5$).}
\end{figure}

\par\end{center}

\subsection{Shock Wave in $r$ \label{subsec:shock-in-r} }

We next consider the shock wave in the variable $r$, which is in
fact a \emph{metric function} (in the sense that $r^{2}=g_{\theta\theta}$).
We use the same numerical setup and same family of timelike geodesics
described in subsection \ref{sub:Nonlinear-Initial-Conditions}. Figure
\ref{fig:r-shock-overview} displays $r(\tau)$ for various timelike
geodesics. The early geodesics shown in panel a reveal the initial
phase of shock formation: While the earliest geodesics look fairly
smooth, the latest ones develop an approximately-vertical section
(clearly seen for e.g. $v_{eh}\ge-1.4$).

Panel b of Fig. \ref{fig:r-shock-overview} presents late geodesics,
demonstrating the nature of the fully-developed shock in $r$. Note
that although a single purple/red curve is visible in this graph,
it actually enfolds many late geodesics in the range $v_{eh}\ge10$.
The dotted black curve displays $r(\tau)$ for an $E=1$ radial geodesic
in pure RN metric with the final asymptotic mass $m_{final}\cong1.4587$
(and $Q=0.95$). The horizontal dashed line marks the corresponding
IH value, $r_{-}\cong0.35177$, of this asymptotic RN metric. The
graph clearly shows that to a very good approximation, all late geodesics
follow the pure-RN geodesic curve --- but only at $r>r_{-}$. When
$r$ approaches $r_{-}$, $r(\tau)$ abruptly decreases, forming the
red vertical section. This is the (fully developed) gravitational
shock wave.

\begin{center}
\begin{figure}[H]
\begin{centering}
\includegraphics[scale=0.25]{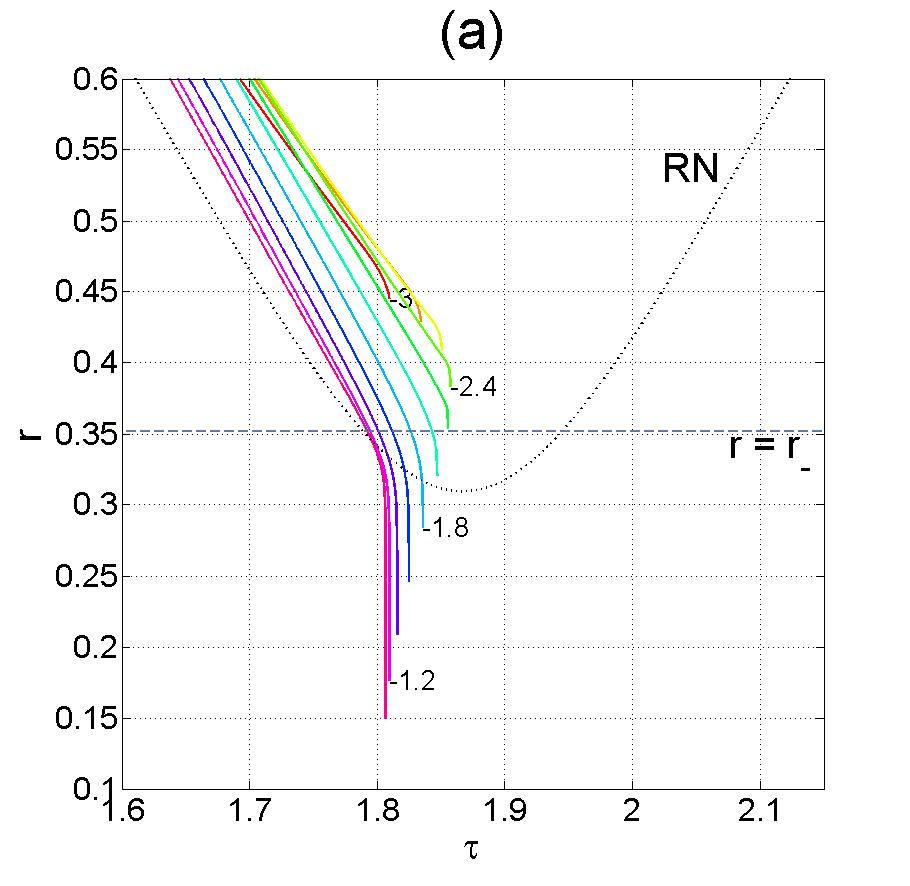}\includegraphics[scale=0.25]{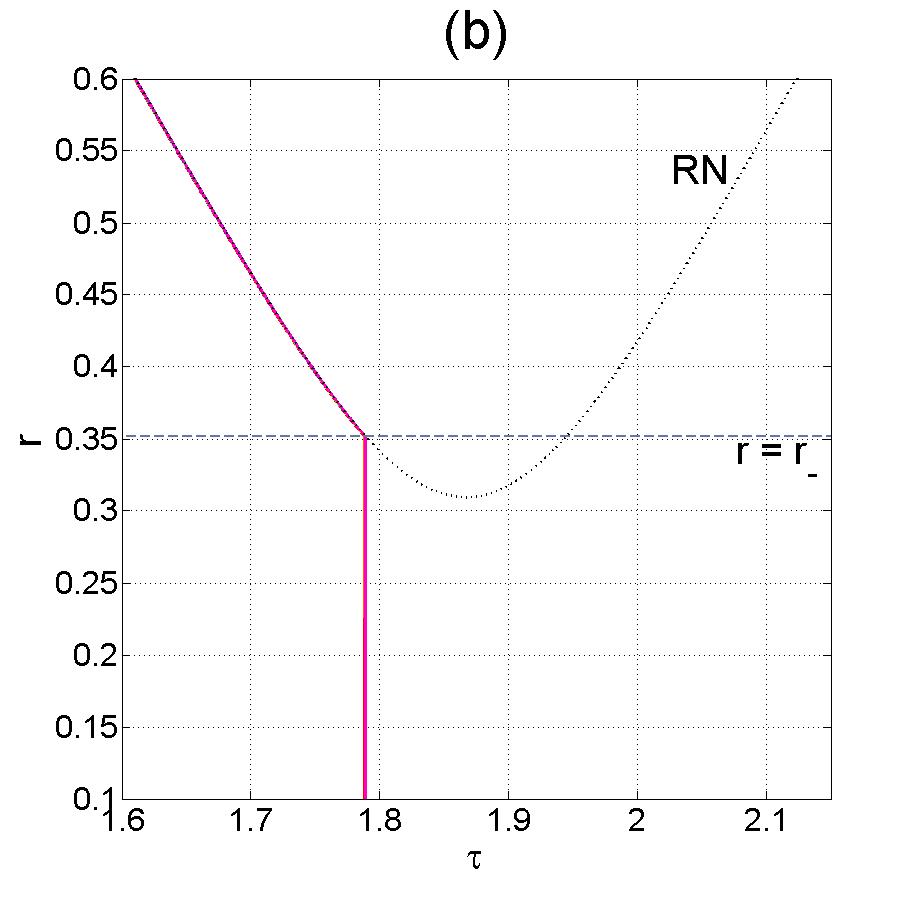}
\par\end{centering}

\protect\caption{\label{fig:r-shock-overview}Numerical results for $r(\tau$) along
the prescribed family of radial timelike geodesics. The black dotted
curve denoted ``RN'' is the theoretical $r(\tau)$ function for
an $E=1$ radial geodesic in the asymptotic RN solution (with parameters
$m=1.45867,\,Q=0.95$); the dashed horizontal line represents the
inner-horizon $r$ value ($r_{-}\protect\cong0.35177$) in that asymptotic
RN. The solid curves represent the numerical results for $r(\tau)$
along the various geodesics. Panel a shows early geodesics (negative
$v_{eh}$ regime, from $-3$ to $-1$). The different geodesics are
marked by different colors and different $v_{eh}$ values; $v_{eh}$
increases by increments of $0.2$. Panel b presents late geodesics
(positive $v_{eh}$ in the range $10\leq v_{eh}\leq80$, with increments
of $5$). In this scale these $15$ different geodesics are unresolved,
they are all represented by the same red/purple curve. (The differences
between these geodesics will be shown in the zoomed figure \ref{fig:r-shock-zoom}.)}
\end{figure}

\par\end{center}

Figure \ref{fig:r-shock-zoom} zooms on the late geodesics near the
``corner'' at $r=r_{-}$. It illustrates how, when $v_{eh}$ increases,
(i) the ``corner'' at each individual geodesic sharpens, and (ii)
the function $r(\tau)$ becomes closer (at $r>r_{-}$) to its asymptotic-RN
counterpart. As $v_{eh}\to\infty$ the limiting function $r(\tau)$
should just coincide with the asymptotic RN curve all the way down
to $r_{-}$, then it should abruptly fall towards $r=0$. 

\begin{center}
\begin{figure}[H]
\begin{centering}
\includegraphics[scale=0.25]{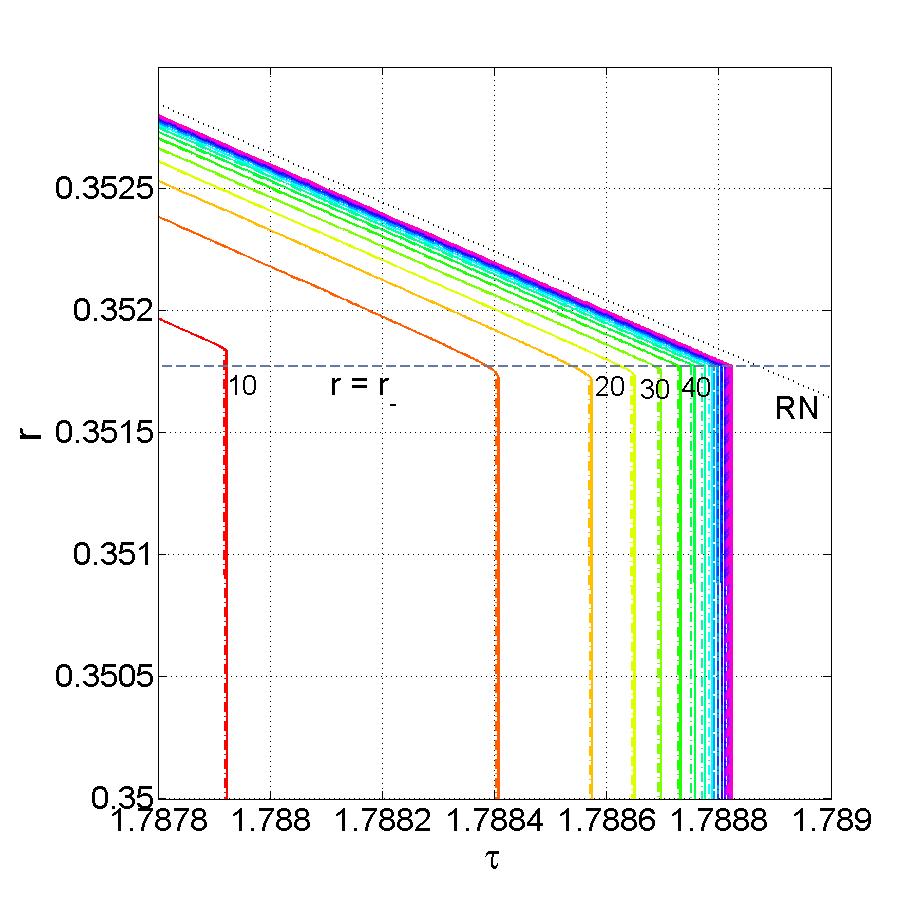}
\par\end{centering}

\protect\caption{\label{fig:r-shock-zoom} Zoom on panel b of figure \ref{fig:r-shock-overview},
near the ``corner''. The figure shows $r(\tau)$ for the same set
of $15$ late geodesics as in Fig. \ref{fig:r-shock-overview}b, but
this time (most of) the individual geodesics are resolved. The different
geodesics are marked by different colors and different $v_{eh}$ values;
$v_{eh}$ increases by increments of $5$ from left ($v_{eh}=10$)
to right ($v_{eh}=80$). Recall that each of the timelike geodesics
is actually represented (here and throughout the paper) by \emph{two}
curves, presenting our two best numerical resolutions $N=640$ (solid
curve) and $N=320$ (dashed curve). In this highly zoomed figure,
these two resolutions can be distinguished, although barely. }
\end{figure}

\par\end{center}

\subsubsection{Shock resolution and shock sharpening}

We again employ a logarithmic presentation (in $\tau-\tau_{final}$)
to resolve the exponentially small proper-time duration of the shock.
Figure \ref{fig:r-shock-log} demonstrates that in this case too,
for late geodesics the function $r(\tau-\tau_{final})$ just shifts
rightward with increasing $v_{eh}$, with no appreciable change in
its form. This shift indicates an exponential decrease in the shock's
proper-time scale. 

To quantify this sharpening, we again need to define a measure $\Delta\tau$
of the shock width (for each geodesic). In this case we cannot take
it to be the proper-time duration between maximum and minimum points
because, unlike the scalar field, $r(\tau)$ is monotonic in the relevant
domain. Instead, we can take $\Delta\tau$ to be the proper time it
takes for $r$ to change from a certain value to another, smaller
one. For concreteness, we define here $\Delta\tau$ to be the proper-time
duration to drop from $0.75\,r_{-}$ to $0.25\,r_{-}$ along the geodesic.
(Here, as before, $r_{-}\cong0.35177$ is the IH radius of the late-time
asymptotic RN geometry.) Figure \ref{fig:r-width-params} displays
this width as a function of $v_{eh}$, confirming again the exponential
sharpening relation $\Delta\tau\propto e^{-\kappa v_{eh}}$ predicted
in Ref. \cite{Marolf-Ori_Shockwave}.

\begin{center}
\begin{figure}[H]
\begin{centering}
\includegraphics[scale=0.25]{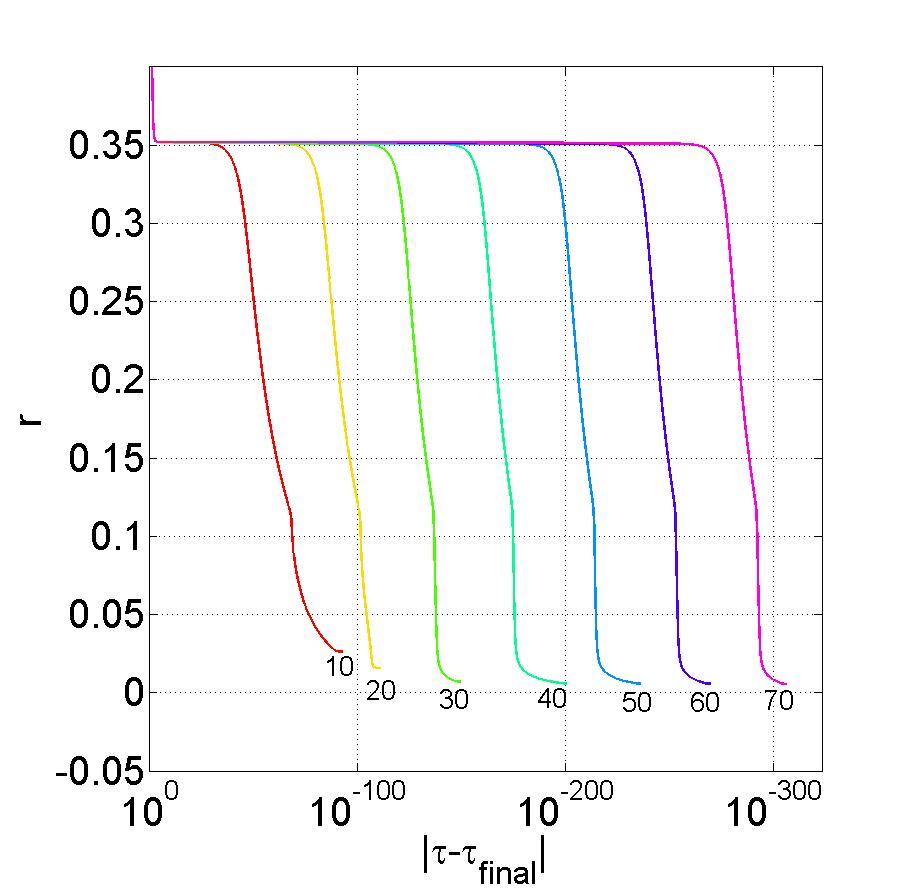}
\par\end{centering}

\protect\caption{\label{fig:r-shock-log}Logarithmic presentation of the numerical
results for $r$ as a function of $|\tau-\tau_{final}|$, along the
prescribed family of radial timelike geodesics. The different geodesics
are marked by different colors and different $v_{eh}$ values; $v_{eh}$
increases by increments of $10$ from left ($v_{eh}=10$) to right
($v_{eh}=70$). }
\end{figure}

\par\end{center}

\begin{center}
\begin{figure}[H]
\begin{centering}
\includegraphics[scale=0.25]{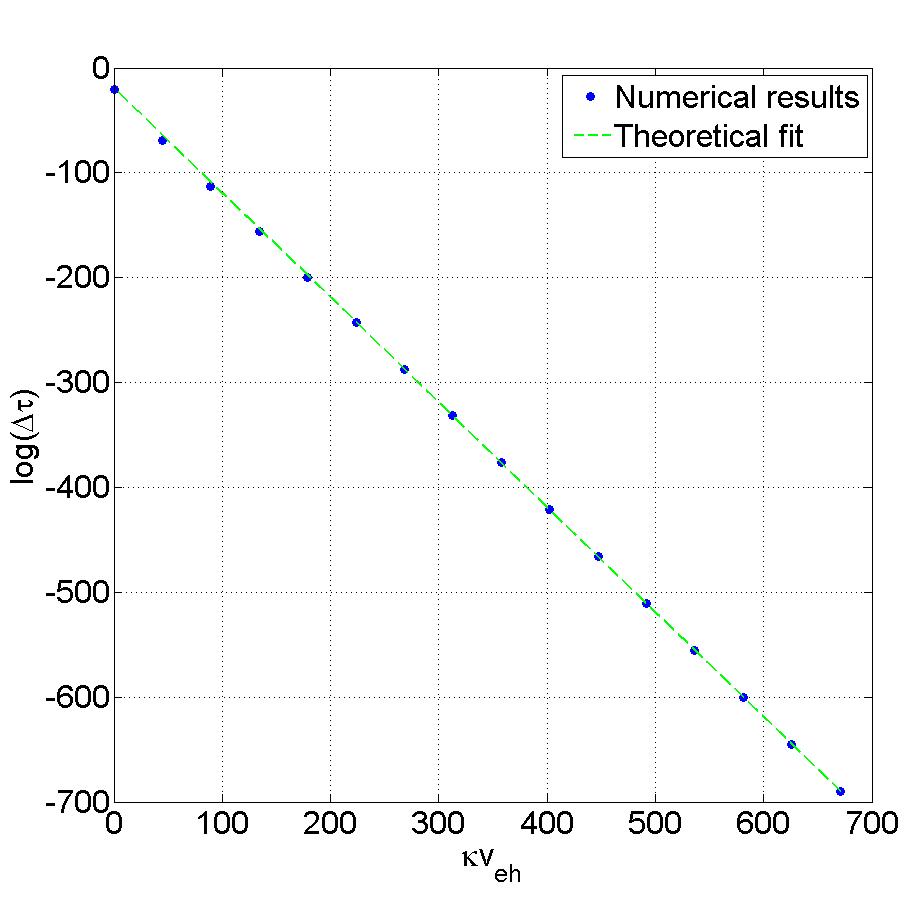}
\par\end{centering}

\protect\caption{\label{fig:r-width-params}Exponential shock sharpening as demonstrated
by the decrease of $\Delta\tau$, the difference between $\tau(r=0.75\,r_{-})$
and $\tau(r=0.25\,r_{-})$. Each point represents a single timelike
radial geodesic; the geodesics are in the range $0\leq v_{eh}\leq75$,
with increments of $5$. The diagonal dashed green line is a fit to
the theoretical relation $\ln(\Delta\tau)=-\kappa v_{eh}+const$.
The graph indicates a very good agreement between the theoretical
prediction and the numerical results, especially for the late-time
geodesics (e.g. $v_{eh}\gtrsim20$). }
\end{figure}

\par\end{center}

\subsection{Probing the shock with ingoing null geodesics}

So far we used timelike geodesics to probe the shock. It is also possible,
however, to use \emph{null} geodesics for that purpose. To this end
we should focus on the dependence of $r$ (or $\Phi$, or any other
variable of interest) on the affine parameter $\lambda$ along a given
null geodesic. The shock will manifest itself as a sharp feature in
$r(\lambda)$ {[}or $\Phi(\lambda)${]}. 

Indeed, timelike geodesics may be considered more ``physical'' than
their null counterparts, as they represent the actual orbits of physical
probes. However, null geodesics are more convenient in several respects:
First of all, the usage of null rather than timelike geodesics reduces
the amount of arbitrariness involved in the construction of the set
of ingoing geodesics, because the degree of freedom of choosing $E$
no longer exists. In particular, if we restrict attention to radial
geodesics (which is obviously the natural thing to do for spherically-symmetric
backgrounds), the family of null geodesics becomes \emph{unique}.
Second, there is no need to solve the geodesic equation in the null
case, because radial null orbits are trivial. In fact, our numerical
results are given on ingoing null geodesics --- namely the grid lines
$v=const$ --- already in the first place. All that is needed, in
order to generate the requested function $r(\lambda)$ {[}or $\Phi(\lambda)${]},
is to compute $\lambda(u)$ along the desired $v=const$ rays. This
is an easy task, as we show in Appendix \ref{sub:Radial-Null}. 

Note that along each null geodesic, $\lambda(u)$ is uniquely defined
up to two free parameters: A global multiplicative constant, and a
global shift. Here we fix both parameters at the EH (for each ingoing
geodesic), by setting $\lambda=0$ and $dr/d\lambda=-1$ at horizon
crossing.

In what follows we shall focus on the shock wave in $r$. The function
$r(\lambda)$ along an ingoing null geodesic admits two convenient
properties: (i) it decreases monotonically, and (ii) it becomes exactly
linear in the pure RN case (see Appendix \ref{sub:The-RN-case}).
Due to these properties, a 3-dimensional graphics turns out to be
especially useful for presenting the shock in $r$. Figure \ref{fig:3d-Null-Geodesic-Formation}
demonstrates the shock formation phase by displaying $r(\lambda)$
for early ingoing null geodesics ($-4<v_{e}<6$), as a function of
their $v_{e}$ value. Notice the smoothness of $r(\lambda)$ for very
early geodesics (say $v_{e}\apprle1$), and the development of an
apparent ``corner'' (at $r\simeq r_{-}$) for larger $v_{e}$. 

Figure \ref{fig:3d-RN-comparison} does the same, but for much later
geodesics (up to $v_{e}\approx85$). In addition, it also displays
the corresponding function $r(\lambda)$ for radial null geodesics
in the exact RN geometry \footnote{This RN geometry is taken with the appropriate asymptotic parameters,
namely $Q=0.95$ and $M=m_{final}$, yielding $r_{+}\simeq2.566$
and $r_{-}\simeq0.352$. } (the straight dotted black lines). This function is given by $r(\lambda)=r_{+}-\lambda$
(independent of $v_{e}$). The 3d graph clearly demonstrates (i) the
nice agreement of $r(\lambda)$ with its RN counterpart up to a clear
``break line'', which occurs at $r\approx r_{-}\simeq0.352$, and
(ii) the sharp, seemingly-vertical, decline of $r(\lambda)$ beyond
that break line. The vertical ``wall'' that forms at $r\apprle r_{-}$
is perhaps the clearest visual presentation of the shock phenomenon. 

\begin{center}
\begin{figure}[H]
\begin{centering}
\includegraphics[scale=0.2]{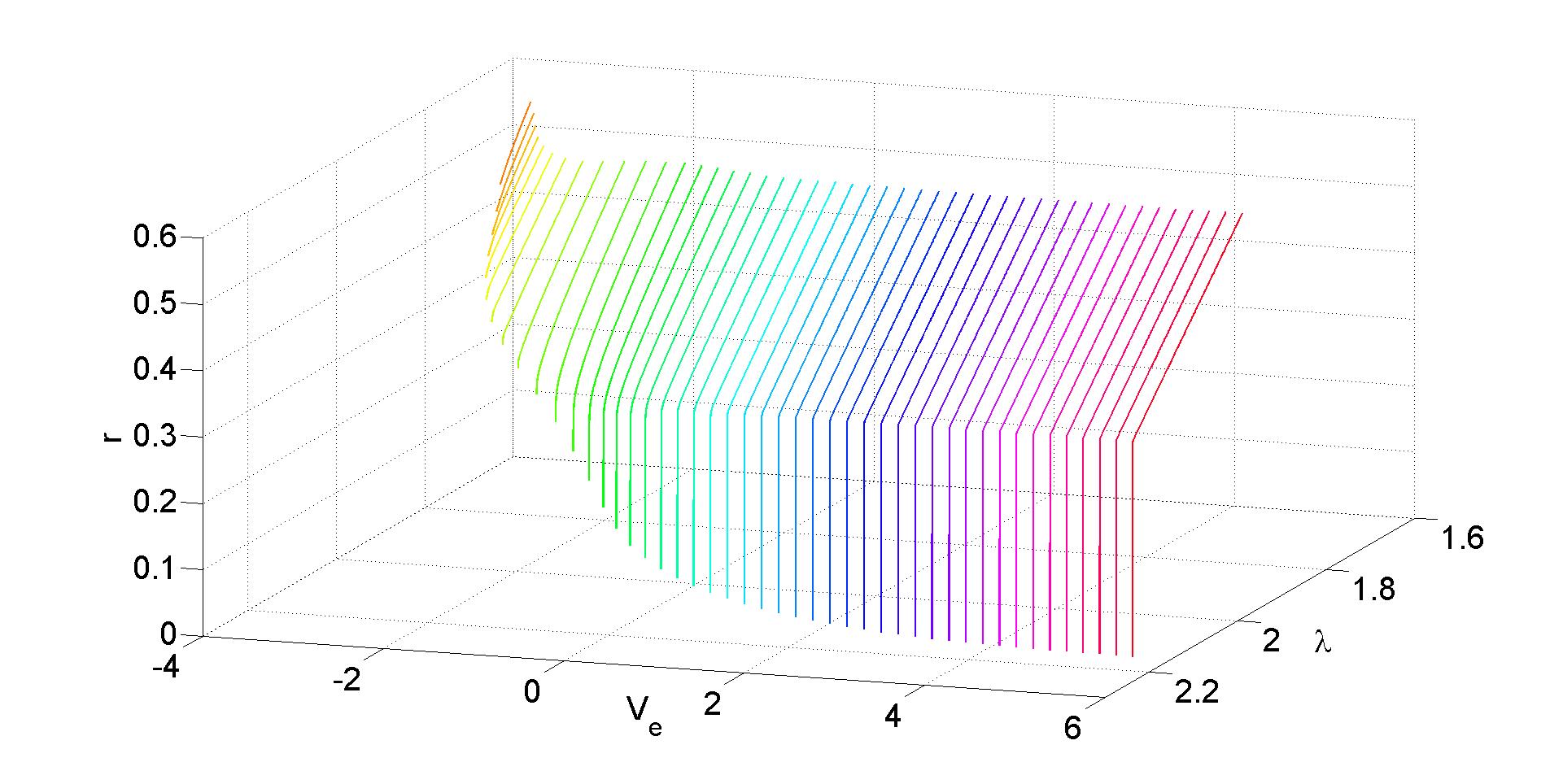}
\par\end{centering}

\protect\caption{\label{fig:3d-Null-Geodesic-Formation}Shock formation in $r$, as
demonstrated by a series of ingoing radial null geodesics in the early
phase $(-4<v_{e}<6)$ of our numerical simulation. Each line represents
$r(\lambda)$ for a single null geodesic (constant $v_{e}$ value).
The gradual development (with increasing $v_{e}$) of a sharp break
point is clearly seen. The location of this break agrees with the
$r$ value of the IH ($r_{-}\protect\cong0.35177$). The geodesics
are displayed here in a limited range of $r$ ($0<r\leq0.6$) in order
to improve shock visibility. (Although, geodesics with very small
$v_{e}$ arrive $u=u_{max}$ at $r$ values well above zero.)}
\end{figure}

\par\end{center}

\begin{center}
\begin{figure}[H]
\begin{centering}
\includegraphics[scale=0.2]{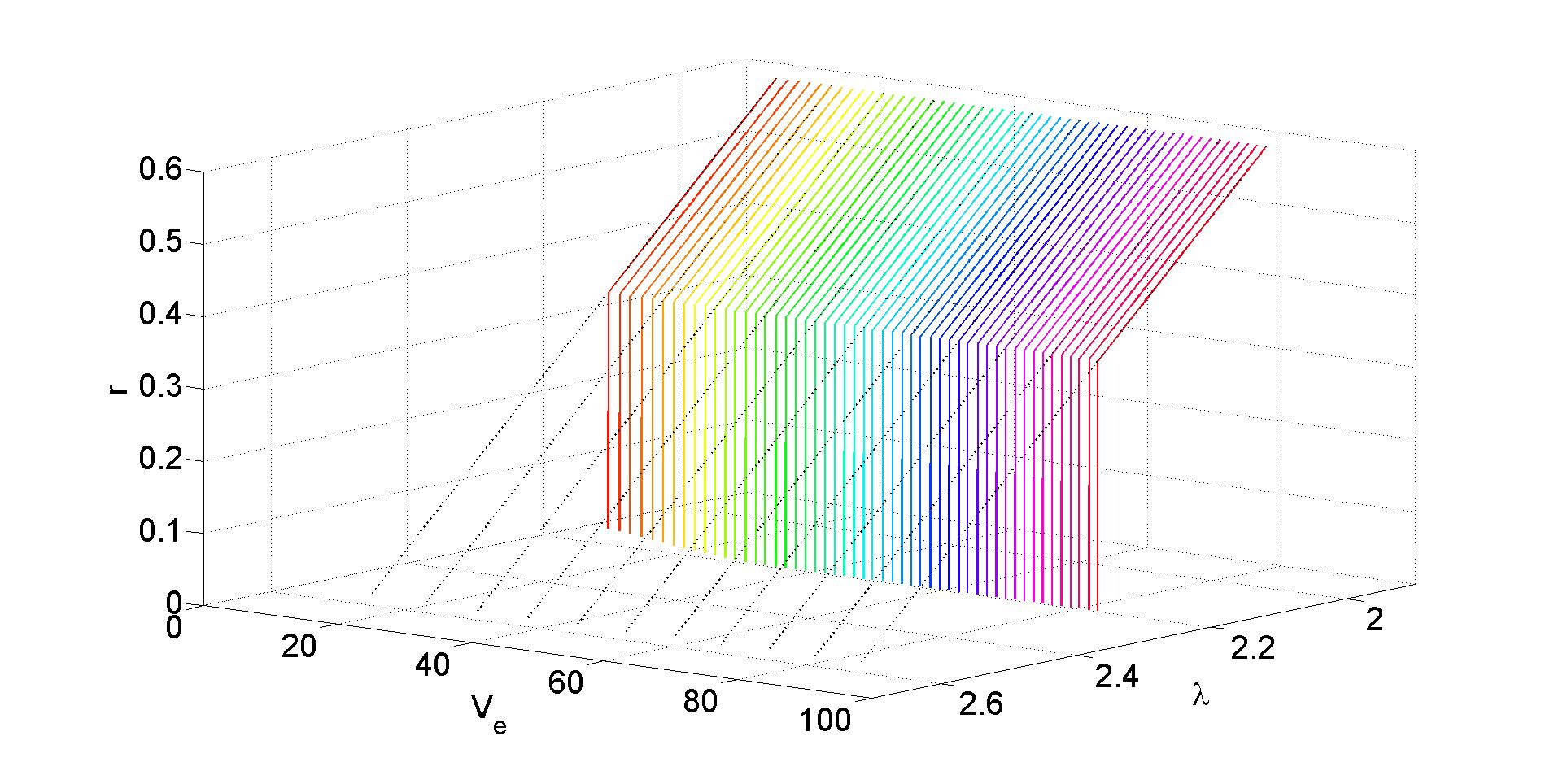}
\par\end{centering}

\protect\caption{\label{fig:3d-RN-comparison}Fully developed shock in $r$, as demonstrated
by a large set of late radial null geodesics (solid colored lines)
in the range $10<v_{e}<85.2$. The figure also displays the corresponding
set of ingoing null geodesics in the exact RN geometry (dotted black
lines). These pure-RN geodesics are given by the linear function $r(\lambda)=r_{+}-\lambda$.
The geodesics are restricted here to a limited range of $r$ ($0<r\leq0.6$)
in order to improve shock visibility. In the range $r\gtrsim r_{-}$
the dotted lines are in fact invisible, because they coincide with
the corresponding solid lines (that is, the geodesics in the perturbed
spacetime are indistinguishable from their pure-RN counterparts).
This situation changes abruptly at the break line at $r\approx r_{-}$.
Beyond that line, at $r\lesssim r_{-}$, the perturbed-spacetime geodesics
suddenly fall (seemingly vertically) towards zero, whereas the RN
geodesics continue their straight linear course. }
\end{figure}

\par\end{center}

\section{DISCUSSION \label{sec:discussion}}

We numerically confirmed the existence of an effective shock at the
left inner-horizon section of a perturbed spherical charged BH, as
predicted by Marolf and Ori (MO) \cite{Marolf-Ori_Shockwave}. We
demonstrated this here for both test scalar-field perturbations and
self-gravitating ones. In both cases, the scalar field $\Phi$ exhibits
an effective shock. In the self-gravitating case, the area coordinate
$r$ also exhibits such an effective shock. Since $r^{2}$ is a metric
function ($=g_{\theta\theta}$), this actually amounts to an (effective)
\emph{gravitational shock wave}. \footnote{Although, in this specific case this gravitational shock is not a
free-gravity phenomenon, it has been triggered by the scalar field.}

This shock expresses itself as an effective discontinuity in $r$
(or similarly $\Phi$) as a function of proper time $\tau$, along
the worldlines of late left-fallers --- namely, free-falling observes
heading toward the left inner-horizon section. As these left-fallers
approach the (would-be) inner horizon, they experience a sudden decrease
in $r$, within an extremely short proper-time interval $\Delta\tau$.
Furthermore, this proper-time scale $\Delta\tau$ decreases exponentially
with the time of infall into the BH. For delay times (from collapse
to jump-in) of typical astrophysical time scales (e.g.) this $\Delta\tau$
becomes tremendously small --- smaller than Planckian by many many
orders of magnitude. 

MO predicted a sharpening rate $\Delta\tau\sim e^{-\kappa v_{eh}}$,
where $\kappa$ is the IH surface gravity of the late-time asymptotic
BH, and $v_{eh}$ is the jump-in time expressed in terms of the asymptotic
Eddington advanced time $v_{e}$. We numerically confirmed this exponential
sharpening rate, for both $r$ and $\Phi$.

The analysis by MO \cite{Marolf-Ori_Shockwave} mostly concentrated
on late-time observers who enter the BH at $v_{eh}\gg M$, after perturbation
tails along the EH has decayed. In their discussion they argued, however,
that earlier observers should also experience the shock (provided
that $v_{eh}\gtrsim$ a few times $M$). We numerically confirmed
this prediction too, as seen in Figs. \ref{fig:test-phi-shock-wave-formation}a,
\ref{fig:nonlinear-phi-shock-wave-formation}, and \ref{fig:r-shock-overview}a.
Our numerical analysis shows the shock development from the early
formation stage (relatively small $v_{eh}$) up to the stage of well-developed
shock (at much larger $v_{eh}$).

In addition to timelike geodesics, we also used null ingoing geodesics
to probe the shock (with $\tau$ replaced by the affine parameter
$\lambda$). This method have some advantages, in particular it allows
a simpler and more transparent presentation of the effective shock
using 3-dimensional graphics. 

We would like to emphasize an important property of the shock: As
was mentioned above, while $v_{eh}$ increases, the shock width $\Delta\tau$
rapidly decreases. However, apart from this sharpening, the shock
amplitude (as well as its internal shape) is insensitive to the increase
in $v_{eh}$. Let's consider the shock in $r$ for concreteness: Very
late left-fallers ($v_{eh}\to\infty$) will experience, enfolded in
the shock, a large decrease in $r$, all the way from $r_{-}$ down
to $r=0$. This is demonstrated in e.g. Fig. \ref{fig:r-shock-log}
or \ref{fig:3d-RN-comparison}. In particular this implies an unbounded,
totally destructive, tidal compression on approaching the (left arm)
IH. This should be contrasted with the experience of late \emph{right-fallers},
heading towards the CH: They experience a bounded tidal deformation
on approaching the CH; and the magnitude of this deformation typically
decreases as an inverse power of $v_{eh}$, and vanishes as $v_{eh}\to\infty$.

This research may be extended in several directions. The most obvious
extension is to consider some other types of perturbations of spherical
charged BHs, and to explore the effective shock formation they induce.
Specifically, the addition of null fluids to our numerical code is
rather straightforward as it merely involves a minor modification
of the constraint equations. MO have not discussed null-fluid perturbation
specifically; however, their arguments could probably be generalized
to this case as well. Note that null fluid is a fairly realistic perturbation
as it can well represent (at least to some extent) the accretion of
cosmic microwave background (CMB) radiation by the BH. Gravitational
and electromagnetic perturbations would also be interesting, but they
are considerably harder to numerically analyze (beyond the linear
level) because of the inevitable breakdown of spherical symmetry. 

The extension of this analysis to \emph{spinning BHs} would be strongly
motivated, because realistic astrophysical BHs are expected to be
spinning. For instance, in both gravitational-wave events GW150914
\cite{GW150914} and GW151226 \cite{GW151226}, the mergers resulted
in BHs with significant spins $a/m\sim0.7$. However, the numerical
study of perturbed spinning BHs would probably be much more challenging,
due to the lack of spherical symmetry. 

In this study, like in Ref. \cite{Marolf-Ori_Shockwave}, we only
considered asymptotically-flat BHs, with perturbations that decay
at late time. Hamilton and Avelino \cite{2010-Hamilton-Avelino} pointed
out, however, that realistic astrophysical BHs steadily accrete dust
as well as CMB photons along cosmological time-scales. It will  be
interesting to extend the shock analysis to such a situation, where
a charged (or spinning) BH accretes for very long times. The null-fluid
formulation will presumably be especially useful for this purpose.

\section*{Acknowledgments}

This research was supported by the Israel Science Foundation (Grant
No. 1346/07).

\appendix

\section{Numerical solution of the geodesic equation\label{sec:NUMERICAL-SOLUTION-OF} }

\subsection{Ingoing Radial Timelike Geodesics\label{sub:Radial-Timelike}}

We shall process here the geodesic equation and bring it to a form
convenient for numerical integration in our numerical code. We begin
with the well known covariant form of the geodesic equation: 
\begin{equation}
\dot{u}_{\alpha}=\frac{1}{2}g_{\mu\nu},_{\alpha}u^{\mu}u^{\nu}\,,\label{eq: covariant_ge_0}
\end{equation}

\noindent where $u_{\alpha}\equiv g_{\alpha\beta}u^{\beta}$, an overdot
denotes $d/d\tau$, and $u^{\mu}\equiv\dot{x}^{\mu}$ (not to be confused
with the null coordinate $u$). Manipulating the right-hand side,
\[
g_{\mu\nu},_{\alpha}u^{\mu}u^{\nu}=g_{\mu\nu},_{\alpha}g^{\mu\varepsilon}u_{\varepsilon}u^{\nu}=-g_{\,\,\,\,,_{\alpha}}^{\mu\varepsilon}g_{\mu\nu}u_{\varepsilon}u^{\nu}=-g_{\,\,\,\,,_{\alpha}}^{\mu\varepsilon}u_{\mu}u_{\varepsilon}\,,
\]
we obtain 
\begin{equation}
\dot{u}_{\alpha}=-\frac{1}{2}g^{\mu\nu},_{\alpha}u_{\mu}u_{v}\,.\label{eq:covariant_ge}
\end{equation}

\noindent Using the line element (\ref{eq:line-element}) we have
$-(1/2)g^{uv}=e^{-\sigma}$, and Eq. (\ref{eq:covariant_ge}) yields

\begin{equation}
\dot{u}_{u}=-2e^{-\sigma}u_{u}u_{v}\sigma,_{u}\:,\;\;\dot{u}_{v}=-2e^{-\sigma}u_{u}u_{v}\sigma,_{v}\,.\label{eq:  udot}
\end{equation}

\noindent We would like to replace this set by another set in which
the independent variable is $u$ (rather than $\tau$). Hereafter
a prime will denote $d/du$ (along the geodesic). Note that $d/du=\left(u^{u}\right)^{-1}d/d\tau$,
or more conveniently, $-(1/2)e^{\sigma}\left(u_{v}\right)^{-1}d/d\tau$.
From Eq. (\ref{eq:  udot}) we now derive the simple equations for
$u_{u}^{'}$ and $u_{v}^{'}$:

\begin{equation}
u_{u}^{'}=\sigma,_{u}u_{u}\:,\;\;u_{v}^{'}=\sigma,_{v}u_{u}\,.\label{eq:geodesic-equations}
\end{equation}

\noindent To complement the set of equations we need the differential
equation for $v(u)$ along the geodesic, namely $v'\equiv dv/du=u^{v}/u^{u}$,
which we rewrite as

\begin{equation}
v^{'}=\frac{u_{u}}{u_{v}}\,.\label{eq:v_of_u}
\end{equation}

Equations (\ref{eq:geodesic-equations}) and (\ref{eq:v_of_u}) form
a closed set of first-order differential equations for the three unknowns
$v(u),u_{u}(u),u_{v}(u)$ --- provided that $\sigma,_{u}$ and $\sigma,_{v}$
are known functions of $u$ and $v$. We use standard predictor-corrector
scheme in order to propagate the unknowns $v,u_{u}$ and $u_{v}$
in $u$ along the geodesics, with finite steps $\Delta u$ (the same
$\Delta u$ parameter that we use in the numerical simulation of the
field equations). The derivatives $\sigma,_{u}$ and $\sigma,_{v}$
are evaluated on the relevant grid points via finite differences,
and then second-order interpolated to the geodesic point $v(u)$.
Similar second-order interpolation is used in order to evaluate the
functions $\Phi,r,\sigma$ (and the mass function $m$) on the geodesic
point once $v(u)$ is known. The proper time $\tau$ along the geodesic
is found via integration of $d\tau/du=1/u^{u}=-(1/2)e^{\sigma}/u_{v}$.

\subsection{Ingoing Radial Null Geodesics\label{sub:Radial-Null}}

We begin with the null analog of the covariant-form geodesic equation
(\ref{eq: covariant_ge_0}):

\begin{equation}
\dot{k}_{\alpha}=\frac{1}{2}g_{\mu\nu},_{\alpha}k^{\mu}k^{\nu}\,,
\end{equation}
where $k^{\mu}\equiv dx^{\mu}/d\lambda$. From the double-null form
of the metric (\ref{eq:line-element}) it immediately follows that
$\dot{k}_{\alpha}=0$ for any radial null geodesic (because for each
such geodesic only one component is non-vanishing, either $k^{u}$
or $k^{v}$). In particular, for our ingoing radial null geodesics,
we get $k_{v}=const$ (and $k_{u}$ vanishes), therefore $du/d\lambda\equiv k^{u}=const\cdot g^{uv}$.
Inverting this relation, and substituting $1/g^{uv}=g_{uv}=-(1/2)e^{\sigma}$,
we obtain $d\lambda/du=Ce^{\sigma}$ where $C$ is an arbitrary normalization
constant. Thus, along each $v=const$ geodesic,

\begin{equation}
\lambda(u)=C\intop^{u}e^{\sigma(\tilde{u},v)}d\tilde{u}\,.\label{eq: A22}
\end{equation}

\noindent Note that this function $\lambda(u)$ is defined up to two
free parameters: the integration constant, and the global normalization
constant $C$.

\paragraph*{\noindent Numerical implementation:}

\noindent The integration is performed retroactively after $\sigma$
is known at all grid points along the geodesic. We use a simple integration
procedure:

\begin{equation}
\lambda(u+\Delta u)=\lambda(u)+C\,e^{[\sigma(u,v)+\sigma(u+\Delta u,v)]/2}\Delta u\,,
\end{equation}
which is second-order accurate. The parameter $C$ is determined by
requiring $d\lambda/dr=-1$ at the EH.

\subsubsection{The RN case \label{sub:The-RN-case}}

Consider now ingoing radial null geodesics in pure RN geometry. When
the latter is expressed in the double-null form (\ref{eq:line-element})
using Eddington coordinates ($u_{e}$,$v_{e}$), one finds $e^{\sigma}=f(r)\equiv1-2M/r+Q^{2}/r^{2}$.
The above expression for $d\lambda/du$ then reduces to $d\lambda/du_{e}=Cf(r)$.
Switching from $u_{e}$ to the tortoise coordinate $r*=(v_{e}-u_{e})/2$
we get $d\lambda/dr*=-2Cf(r)$ (recalling that $v_{e}$ is constant
along the ingoing ray). Since $r*$ satisfies $dr/dr*=f(r)$, we find
that $d\lambda/dr=-2C$. Thus, $\lambda(r)$ is linear along ingoing
null geodesics (and the same for outgoing ones). 

Again we have two arbitrary constants ($C$ and the integration constant),
and as before we choose them such that at the EH $\lambda$ vanishes
and $d\lambda/dr=-1$, obtaining $r(\lambda)=r_{+}-\lambda$.


\begin{thebibliography}{10}
\bibitem{1968-Penrose-PRN-Analytical}R. Penrose, \textquotedbl{}Structure
of Space-Time'' in Battelle Rencontres, edited by C. de Witt and
J. Wheeler (W. A. Benjamin, New York, 1968), p. 222.

\bibitem{1973-Simpson-Penrose-PRN-Numerical}M. Simpson and R. Penrose,
\textquotedbl{}Internal Instability in a Reissner-Nordström Black
Hole\textquotedbl{}, Int. J. Theor. Phys. \textbf{7}, 183 (1973).

\bibitem{1981-Hiscock-PRN-Analytical}W. A. Hiscock, \textquotedbl{}Evolution
of the interior of a charged black hole\textquotedbl{}, Phys. Lett.
\textbf{83A}, 110 (1981).

\bibitem{Novikov}Y. Gursel, I. D. Novikov, V. D. Sandberg, and A.
A. Starobinsky, \textquotedbl{}Final state of the evolution of the
interior of a charged black hole\textquotedbl{}, Phys. Rev. D \textbf{20},
1260 (1979).

\bibitem{1982-Chand-Hartle-PRN-Analytical}S. Chandrasekhar and J.
B. Hartle, \textquotedbl{}On crossing the Cauchy horizon of a Reissner-Nordström
black-hole\textquotedbl{}, Proc. R. Soc. London \textbf{A384}, 301
(1982).

\bibitem{Poisson-Israel-PRL}E. Poisson and W. Israel, ``Inner-horizon
instability and mass inflation in black holes'', Phys. Rev. Lett.
\textbf{63}, 1663 (1989).

\bibitem{1990-Poisson-Israel-mass-function}E. Poisson and W. Israel,
\textquotedbl{}Internal structure of black holes\textquotedbl{}, Phys.
Rev. D \textbf{41}, 1796 (1990).

\bibitem{Ori-Weak-singularity}A. Ori, \textquotedbl{}Inner structure
of a charged black hole - an exact mass-inflation solution\textquotedbl{},
Phys. Rev. Lett. \textbf{67}, 789 (1991).

\bibitem{1993-Gnedin}M. L. Gnedin and N. Y. Gnedin, \textquotedbl{}Destruction
of the Cauchy horizon in the Reissner-Nordström black hole\textquotedbl{},
Class. Quantum Grav. \textbf{10}, 1083 (1993).

\bibitem{1995-Brady-Smith}P. R. Brady and J. D. Smith, \textquotedbl{}Black
Hole Singularities: A Numerical Approach\textquotedbl{}, Phys. Rev.
Lett., \textbf{75}, 1256 (1995).

\bibitem{1997-Burko-Internal-structure}L. M. Burko, \textquotedbl{}Structure
of the Black Hole's Cauchy-Horizon Singularity\textquotedbl{}, Phys.
Rev. Lett. \textbf{79}, 4958 (1997) .

\bibitem{1998-Hod-Piran}S. Hod and T. Piran, \textquotedbl{}Mass
Inflation in Dynamical Gravitational Collapse of a Charged Scalar
Field\textquotedbl{}, Phys. Rev. Lett. \textbf{81}, 1554 (1998). 

\bibitem{Burko-EOMs}See e.g. L. M. Burko and A. Ori, ``Analytic
study of the null singularity inside spherical charged black holes'',
Phys. Rev. D \textbf{57}, R7084 (1998).

\bibitem{Tipler}F. J. Tipler, \textquotedbl{}Singularities in conformally
at spacetimes\textquotedbl{}, Phys. Lett. \textbf{64A}, 8 (1977).

\bibitem{Ori-What-is-Weak}A. Ori, \textquotedbl{}Strength of curvature
singularities\textquotedbl{}, Phys. Rev. D \textbf{61}, 064016 (2000).

\bibitem{Numerical-Methods} E. Eilon and A. Ori, ``Adaptive gauge
method for long-time double-null simulations of spherical black-hole
spacetimes'', Phys. Rev. D \textbf{93}, 024016 (2016).

\bibitem{Dafermos}M. Dafermos, \textquotedbl{}Stability and Instability
of the Cauchy Horizon for the Spherically Symmetric Einstein-Maxwell-Scalar
Field Equations\textquotedbl{}, Ann. Math. \textbf{158}, 875 (2003).

\bibitem{1992-Ori-spinning}A. Ori, ``Structure of the singularity
inside a realistic rotating black hole'', Phys. Rev. Lett. \textbf{68},
2117 (1992).

\bibitem{Brady-Droz}P. R. Brady, S. Droz, and S. M. Morsink, ``Late-time
singularity inside nonspherical black holes'', Phys. Rev. D \textbf{58},
084034 (1998).

\bibitem{2000-Ori-Spinning}A. Ori, ``Oscillatory Null Singularity
inside Realistic Spinning Black Holes'', Phys. Rev. Lett. \textbf{83},
5423 (1999). 

\bibitem{2010-Hamilton-Avelino}For a somewhat different view-point
see, however, A. J. S. Hamilton and P. P. Avelino, \textquotedbl{}The
physics of the relativistic counterstreaming instability that drives
mass inflation inside black holes\textquotedbl{}, Phys. Rep. 495,
1 (2010).

\bibitem{Marolf-Ori_Shockwave}D. Marolf and A. Ori, ``Outgoing gravitational
shock wave at the inner horizon: The late-time limit of black hole
interiors'', Phys. Rev. D \textbf{86}, 124026 (2012).

\bibitem{Marolf-Extreme}This study by MO \cite{Marolf-Ori_Shockwave}
was triggered by the picture of extreme black holes at late times
suggested in D. Marolf, \textquotedblleft The dangers of extremes,\textquotedblright{}
Gen. Rel. Grav. 42, 2337 (2010). {[}arXiv:1005.2999 {[}gr-qc{]}{]}.

\bibitem{GW150914}P. Abbott et al. (LIGO Scientific and Virgo Collaborations),
``Observation of Gravitational Waves from a Binary Black Hole Merger'',
Phys. Rev. Lett. \textbf{116}, 061102 (2016). 

\bibitem{GW151226}P. Abbott et al. (Virgo and LIGO Scientific Collaborations),
``GW151226: Observation of Gravitational Waves from a 22-Solar-Mass
Binary Black Hole Coalescence'', Phys. Rev. Lett. \textbf{116}, 241103
(2016) .\end{thebibliography}
\end{document}